\documentclass[sigconf]{acmart}

\copyrightyear{2026}
\acmYear{2026}
\acmConference[WWW '26] {Proceedings of the ACM Web Conference 2026}{April 13--17, 2026}{Dubai, United Arab Emirates.}
\acmBooktitle{Proceedings of the ACM Web Conference 2026 (WWW '26), April 13--17, 2026, Dubai, United Arab Emirates}

\newcommand{\name}{VarParser\xspace}

\usepackage{amsmath,amssymb,amsfonts}
\usepackage{textcomp}
\usepackage{xcolor}
\usepackage{algorithmic}
\usepackage{algorithm}
\usepackage{balance}
\usepackage{graphicx}
\usepackage{booktabs}
\usepackage{colortbl}
\usepackage{balance}
\usepackage{makecell}
\usepackage{multirow}
\usepackage{array}
\usepackage{color}
\usepackage{float}
\usepackage{xspace}
\usepackage{enumitem}
\usepackage{microtype}

\usepackage{changes}

\begin{document}


\title{VarParser: Unleashing the Neglected Power of Variables for LLM-based Log Parsing}
\author{Jinrui Sun}
\affiliation{%
  \institution{Peking University}
  \city{Beijing}
  \country{China}
  \postcode{43017-6221}
}
\email{jrsun25@stu.pku.edu.cn}

\author{Tong Jia}
\affiliation{%
  \institution{Peking University}
  \city{Beijing}
  \country{China}
}
\authornote{Corresponding author.}
\email{jia.tong@pku.edu.cn}

\author{Minghua He}
\affiliation{%
  \institution{Peking University}
  \city{Beijing}
  \country{China}
}
\email{hemh2120@stu.pku.edu.cn}

\author{Ying Li}
\affiliation{%
  \institution{Peking University}
  \city{Beijing}
  \country{China}
}
\email{li.ying@pku.edu.cn}

\renewcommand{\shortauthors}{Jinrui Sun, Tong Jia, Minghua He, and Ying Li.}

\begin{abstract}

Logs serve as a primary source of information for engineers to diagnose failures in large-scale online service systems. Log parsing, which extracts structured events from massive unstructured log data, is a critical first step for downstream tasks like anomaly detection and failure diagnosis. With advances in large language models (LLMs), leveraging their strong text understanding capabilities has proven effective for accurate log parsing. However, existing LLM-based log parsers all focus on the constant part of logs, ignoring the potential contribution of the variable part to log parsing. This constant-centric strategy brings four key problems. First, inefficient log grouping and sampling with only constant information. Second, a relatively large number of LLM invocations due to constant-based cache, leading to low log parsing accuracy and efficiency. Third, a relatively large number of consumed constant tokens in prompts leads to high LLM invocation costs. At last, these methods only retain placeholders in the results, losing the system visibility brought by variable information in logs.

Facing these problems, we propose a variable-centric log parsing strategy named \name. Through variable contribution sampling, variable-centric parsing cache, and adaptive variable-aware in-context learning, our approach can efficiently capture the variable parts of logs and leverage their contributions to parsing. By introducing variable units, we preserve rich variable information, enhancing the integrity of log parsing results. Extensive evaluations on large-scale datasets demonstrate that \name achieves higher accuracy compared to existing methods, significantly improving parsing efficiency while reducing the LLM invocation costs.
\end{abstract}



\maketitle

\section{Introduction}
\label{sec:intro}

\begin{figure}[htp]
    \centering
    \includegraphics[width=\linewidth]{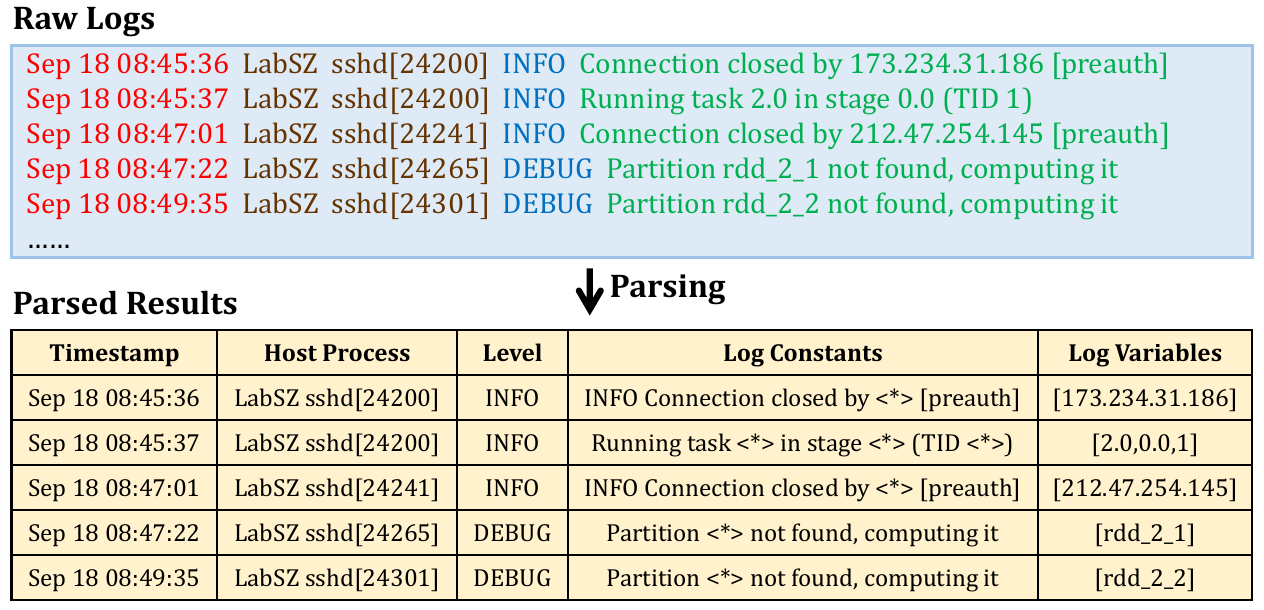}
    \caption{An example of log parsing procedure. The red part of raw logs represents timestamps, the brown part represents the host and process information, the blue part represents the log level, and the green part represents the log content.}
    \label{fig:parse}
\end{figure}

In recent years, online services such as Google Search, Bing, Facebook, and Twitter have grown rapidly, providing 24/7 support to millions of users. Despite significant efforts to ensure service reliability and availability, hardware and software failures remain inevitable in practice, often leading to unplanned outages. To address these incidents, practitioners heavily rely on system logs as a key diagnostic resource. These logs contain rich historical records of events and system states, enabling effective anomaly detection, failure diagnosis, and recovery support. However, with the rapid growth in raw log volume, identifying valuable information from massive log data has become increasingly challenging, even for experienced engineers \cite{onion}. To address this issue, automated log analysis techniques have emerged in recent years. A critical and widely adopted first step in this process is log parsing, which involves converting semi-structured console logs into a structured format. This conversion serves as a crucial preliminary step for automated log analysis tasks such as anomaly detection \cite{DeepLog, MidLog, MetaLog, HumanLog, hilogx}, root cause analysis \cite{logflatter, LogRule, CloudRCA, Toomanycooks}, and failure diagnosis \cite{Medicine, cloud19, logcluster, LogSed, LogFlash}, etc. Typically, raw log messages consist of two parts: (1) Log constants, which represent the constant part describing the main content of the recorded event; (2) Log variables, which contain the dynamic part associated with the event (determined at runtime). Figure \ref{fig:parse} shows the procedure of log parsing.

Existing log parsing methods can be categorized into three types: Syntax-based parsers, Semantic-based parsers, and LLM-based parsers. Syntax-based parsers \cite{Drain, logcluster-row, Hue, Brain} extract templates using strategies such as frequency-based, similarity-based, and heuristic methods, which often struggle to accurately identify templates when log messages deviate from manually defined rules. Semantic-based parsers rely on neural networks to differentiate templates from variables but require large amounts of labeled data for training. LLM-based parsers leverage pre-trained large language models and optimize template extraction through in-context learning (ICL). Many studies have shown that, compared to syntax-based and semantic parsers, LLM-based parsers achieve significantly higher accuracy \cite{LLMParser, LILAC, logparser-llm, Lunar, LibreLog}. LLM-based log parsing methods typically involve three steps: (1) log sampling, where a candidate sampling algorithm selects a small set of diverse and information-rich historical logs to provide effective demonstrations for the LLM; (2) log matching, which matches the already parsed templates with the new logs to avoid redundant LLM invocations; (3) in-context learning, which allows the LLM to learn domain-specific knowledge and generate log templates based on provided log examples. 

Existing LLM-based log parsers all focus on the constants of logs, ignoring the potential contribution of variable information in the original logs to log parsing. This constant-centric strategy suffers from four key problems. First, inefficient log grouping and sampling with only constant information. Second, a relatively large number of LLM invocations due to constant-based cache, leading to low log parsing efficiency. Third, a relatively large number of consumed constant tokens in prompts leads to high LLM invocation costs. Lastly, these methods only retain placeholders in the results, losing the system visibility brought by variable information in logs. The specific approach of the existing method \cite{DivLog, logbatcher, LibreLog, Lunar, logparser-llm, LILAC} is constant-centric in the following way: 

\begin{itemize}[leftmargin=*]
    \item \textbf{In log sampling}: Existing methods are dominated by the constants in logs for grouping criteria (e.g., top-k frequent tokens, log similarity, etc.), aiming to group logs with the same constants together. This results in the number of log groups being approximately equal to the number of templates, thereby limiting sampling efficiency.
    \item \textbf{In log matching}: Existing methods only store the constants in cache and fail to capture dynamically variable positions, supporting only token matching, which leads to a high frequency of matching failures.
    \item \textbf{In in-context learning}: Existing methods select log-template pairs as examples based on the similarity of log levels. Since constants typically account for a large proportion of logs, the constant part often dominates the selection of examples. However, as LLMs can understand natural language, the repeated natural language in the examples contributes little to understanding the new logs to be parsed, resulting in token wastage and higher LLM invocation costs.
    \item \textbf{Result integrity}: Existing log parsing methods only preserve log templates (constants + placeholders), discarding all variable information during parsing. However, log variables often carry critical content such as system states, internal details, and encoded values. Losing this information greatly reduces system visibility and weakens the effectiveness of downstream tasks.
\end{itemize}


Facing these problems, we demonstrate a new log parsing strategy called the variable-centric strategy. We propose \name, a \underline{var}iable-centric log \underline{parser} with large language models. We focus on the variable part of the logs throughout parsing to fully leverage the contribution of the variables to the log parsing process. Specifically, our approach is variable-centric in the following way: 

\begin{itemize}[leftmargin=*]
    \item \textbf{In log sampling}: \name clusters uncommon tokens to group historical logs with the same variables together and introduces a log contribution scoring and sampling method based on variable distributions. Since logs with the same variables cover a broader scope, the number of log groups sampled through this method is much smaller than the number of templates, thereby improving sampling efficiency.
    \item \textbf{In log matching}: \name proposes variable-fuzzed matching table and incorporates variable units into the cache, and proposes mask matching that considers cached variable information. This enables more comprehensive log matching, reduces the number of LLM invocations, and improves parsing efficiency.
    \item \textbf{In in-context learning}: \name adaptively selects token-level variable information as examples, capturing finer-grained information while significantly reducing token consumption.
    \item \textbf{Result integrity}: By caching variable units, \name enables the extraction of variable types, specific instances (including counts and examples), and their occurrence frequencies across the entire log set. This variable-related information offers engineers enhanced system visibility.
\end{itemize}


The comprehensive experiments on the public large-scale log dataset Loghub-2.0\cite{loghub-2.0} show that \name achieves the highest average accuracy across all performance metrics, significantly improving parsing efficiency compared to the state-of-the-art baseline LLM-based parser and reducing the LLM invocation costs. 

Unlike prior work \cite{vista} that simply explored the feasibility of efficiency, \name performs large-scale experiments and introduces variable-fuzzed matching and a redesigned caching and matching mechanism, achieving the effectiveness and efficiency of variable-centric log parsing. The main contributions of this work are summarized as follows:

\begin{itemize}[leftmargin=*]
    \item We present a new log parsing strategy called the variable-centric strategy and propose the first variable-centric LLM-based log parser, named \name, which fully leverages the neglected power of variable information in the raw logs to log parsing.
    \item By caching variable units, \name preserves variable labels and instances, retaining richer variable-related information and significantly improving result integrity.
    \item We evaluate \name on large-scale public datasets. The results show that \name outperforms all existing log parsing methods in accuracy, significantly improving the parsing efficiency of LLM-based log parsers while reducing the overall LLM invocation costs.
\end{itemize}


\section{Background and Related Work}

\subsection{Syntax-based Log Parsers}

Syntax-based log parsers extract templates by identifying recurring patterns in logs, treating them as constants and other elements as variables. These methods include frequency-based, similarity-based, and heuristic approaches. For instance, Drain \cite{Drain} employs a fixed-depth prefix tree structure to effectively extract common templates using specific heuristics, such as prefix tokens and log length. However, these methods are constrained by predefined rules, and their parsing accuracy drops significantly when log data deviates from these rules.

\subsection{Semantic-based Log Parsers}
Semantic-based log parsers typically use deep learning models to capture the semantics within log messages. For instance, Uniparser \cite{Uniparser} formulates log parsing as a token classification problem and uses a Bi-LSTM \cite{bi-lstm} model for training, while VALB \cite{VALB} treats it as a sequence labeling problem and employs a Bi-LSTM-CRF \cite{bi-lstm-crf} model. However, these approaches depend heavily on large amounts of labeled data, and their performance is severely limited when labeled data is scarce or log formats change frequently. This limitation can cause significant performance drops when dealing with complex and large-scale log data. 

\subsection{LLM-based Log Parsers}
\label{sec:llm_parsers}

Large language models demonstrate strong capabilities in understanding programming languages, showing great potential for software engineering applications. However, effectively applying LLMs to domain-specific tasks remains challenging. While fine-tuning on specific datasets can yield good performance, it is computationally expensive and requires large amounts of high-quality labeled data \cite{codet5}. This process is also time-consuming and heavily reliant on data availability and quality, making it impractical in data-scarce scenarios. To overcome these limitations, in-context learning has emerged as a key technique, enabling LLMs to adapt to new tasks without explicit retraining \cite{ICL-survey1, ICL-survey2}. By leveraging examples within prompts, ICL allows models to learn from contextual cues and has proven effective in domain-specific downstream tasks \cite{ICL-medi-1, ICL-sensi-1}. As a result, recent studies have adopted ICL to develop LLM-based log parsers \cite{DivLog, logparser-llm, LILAC, Lunar, logbatcher}, which achieve significantly higher accuracy than traditional syntax- or semantics-based methods. However, as noted in Section \ref{sec:intro}, these approaches focus primarily on constants in log sampling, parsing cache, and in-context learning, leading to inefficiency, high LLM invocation costs, and compromised parsing integrity.

\section{Methodology}

\begin{figure*}[t]
    \centering
    \includegraphics[width=\linewidth]{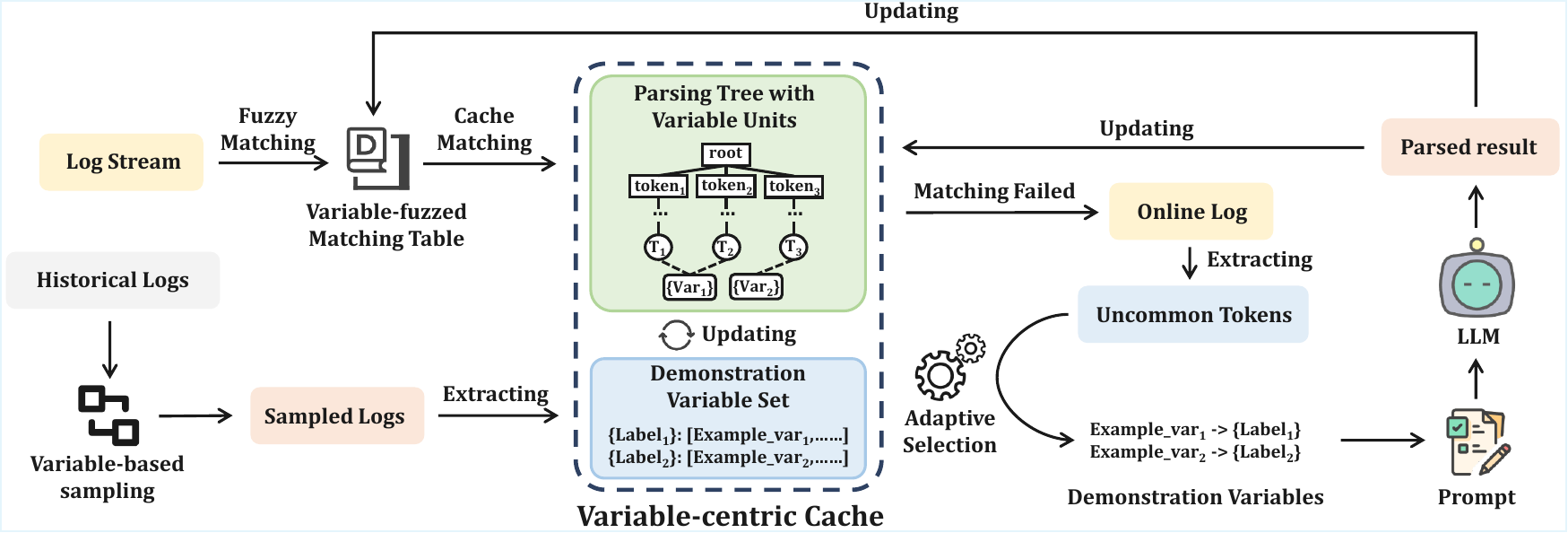}
    \caption{The overall framework of \name. The demonstration variable set is initially extracted from historical logs through dynamic contribution sampling. The incoming log stream is matched against the cache. When matching fails, uncommon tokens are leveraged to select demonstration variables for LLM prompting.}
    \label{fig:overview}
\end{figure*}

In this section, we propose \name, the first variable-centric general-purpose log parsing method based on large language models. \name introduces three core modules: Variable contribution sampling, Variable-centric parsing cache, and Adaptive Variable-aware in-context learning. Following the pipeline of previous work \cite{logparser-llm, LibreLog, LILAC}, \name first extract a small set of diverse and information-rich log messages and manually label them from historical logs. For the online log stream, our method sequentially parses them one by one. The overall framework of \name is shown in Figure \ref{fig:overview}.

\subsection{Variable Contribution Sampling}

To extract a small, diverse, and information-rich set of log messages from a large volume of log data, we propose variable contribution sampling, which consists of three components: Grouping, Contribution Calculating, and Greedy Diversified Sampling. 

\begin{figure}[htp]
    \centering
    \includegraphics[width=\linewidth]{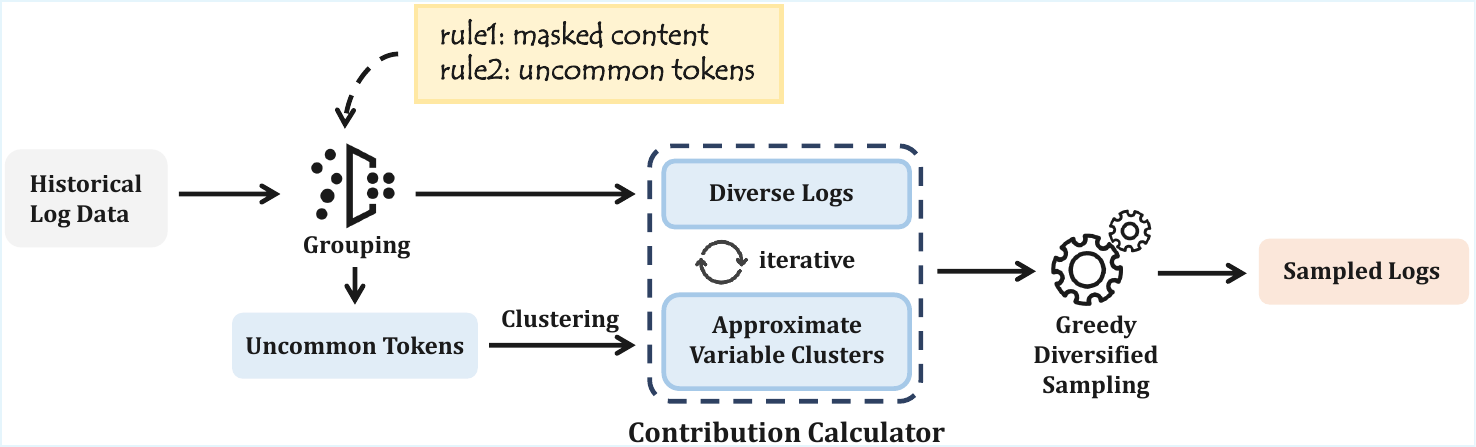}
    \caption{The workflow of variable contribution sampling, where both variable contribution and log contribution are dynamically updated during sampling.}
    \label{fig:sample}
\end{figure}

\subsubsection{\textbf{Grouping}} 

During the log sampling process, redundant log data can significantly increase the complexity and computational cost of data processing. Therefore, eliminating duplicate historical logs is essential to minimizing the size of the sampling pool. To achieve this, we start by masking common dynamic content in the logs (e.g., numbers, IP addresses) to perform preliminary grouping.

Since LLM can understand natural language, the repeated natural language (i.e., constants) included in the examples contribute very little to the understanding of new logs to be parsed. Logs from different templates containing the same type of variables provide roughly equivalent domain-specific knowledge to the LLM. In other words, logs with identical variables, regardless of their template, contribute similarly to the log parsing process. For this purpose, we first tokenize each log into a list of tokens using a delimiter. We then use common words from the NLTK toolkit \cite{nltk} to filter out a set of uncommon tokens for each log. The rationale behind this approach is that log variables typically do not appear in lists of common words. Finally, we group the logs based on these uncommon tokens, ensuring that each log in the sampling pool contains unique variable information.

\subsubsection{\textbf{Contribution Calculating}} 

To evaluate the contribution of each log message to the LLM's parsing of new logs, we first cluster all uncommon tokens to group tokens that belong to the same type of variable. In this method, we mask the digit parts of each token to focus on the non-numeric content, and then use the Jaccard similarity to measure the similarity between tokens. Tokens with a similarity score greater than the threshold are grouped into the same cluster, effectively forming approximate variable clusters. Specifically, we treat each token as a set of characters and calculate the Jaccard similarity based on the intersection and union of these character sets. The formula for Jaccard similarity is $JS = \frac{A \cap B}{A \cup B}$. Tokens with a higher similarity score are considered to belong to the same variable type and are thus clustered together. 

Next, we calculate the contribution of each approximate variable cluster. If a cluster appears frequently across different logs, it indicates greater information potential, and its contribution to the parsing of new logs is accordingly greater. Specifically, we quantify the contribution of a cluster by the number of distinct positions it occupies across logs. As illustrated in Figure \ref{fig:cluster}, cluster C1 contributes more significantly than cluster C2 due to its presence across more different logs. Finally, we compute the contribution of each log by aggregating the contributions of all approximate variable clusters within that log. 

\subsubsection{\textbf{Greedy Diversified Sampling}}

To make the sampling results more information-rich and diverse, we propose Greedy Diversified Sampling. This algorithm prioritizes sampling the log with the highest contribution and dynamically updates the contribution of each log during the sampling process. Specifically, after sampling a log with the highest contribution, the contribution of all the approximate variable clusters contained in that log is reset to zero, and the contribution of each log is then updated before repeating the process. The comprehensive sampling rule is elucidated in Algorithm \ref{alg: gds}. This algorithm allows unsampled logs to dynamically adjust their priority based on their remaining information potential.

\begin{algorithm}[htp]
    \caption{Greedy Diversified Sampling}
    \label{alg: gds}
    \renewcommand{\algorithmicrequire}{\textbf{Input:}}
    \renewcommand{\algorithmicensure}{\textbf{Output:}}
    
    \begin{algorithmic}[1]
        \REQUIRE $logs, var\_contributions$     
        \ENSURE $sampled\_logs$    
        
        \STATE $sampled\_tokens \leftarrow \{\}$ 
        \STATE $sampled\_logs \leftarrow \{\}$
        \STATE $log\_contributions = calculate(var\_contributions)$
        
        \WHILE{$len(sampled\_logs) < sampled\_num$}
            \STATE $new\_log \leftarrow select(logs, log\_contributions)$
            \STATE $sampled\_logs.add(new\_log)$
            \STATE $uncommon\_tokens \leftarrow filter\_token(new\_log)$
        
            \FOR{$token \in uncommon\_tokens$}
                \STATE $var \leftarrow get\_cluster(token)$
                \STATE $var\_contributions[var] = 0$
            \ENDFOR
    
            \STATE $log\_contributions = calculate(var\_contributions)$
        \ENDWHILE
        \RETURN $sampled\_logs$
        
    \end{algorithmic}
\end{algorithm}

\subsection{Variable-Centric Parsing Cache}

Following previous work \cite{Drain, LILAC, logparser-llm}, we use a prefix tree structure to store templates generated by the LLM, serving as a parsing cache. In this cache, all generated log templates are tokenized into token lists and stored sequentially in the tree. Each intermediate node in the tree represents a token, where \verb|"<*>"| acts as a wildcard token that can match tokens of any length. Each leaf node represents a unique log template, corresponding to the concatenation of all tokens along the unique path from the root to the leaf. We optimized the cache by introducing the concept of variable units into the parsing cache. Each variable unit is assigned a label that represents its information. Specifically, for every parsed log, its variable components are stored in their respective variable units based on their labels, while templates are associated with these variable units along with corresponding positional information.

\subsubsection{\textbf{Variable-fuzzed Matching Table}}

To improve parsing accuracy and efficiency, we first use mask common dynamic content (e.g., numbers, file path, IP addresses) in logs through regular expressions to create variable-fuzzed content, then cache a variable-fuzzed matching table that maps each log's fuzzed content to its parsed template. When a new log is provided, \name queries this table to check if its template is already in the cache, reducing redundant LLM queries and improving efficiency. Table \ref{tab:fuzzy} demonstrates the variable fuzzing strategy, where all common dynamic content is replaced with "\{Variable\}", and consecutive occurrences of "\{Variable\}" are merged.

\subsubsection{\textbf{Cache matching and updating}} 

\begin{figure}[htp]
    \centering
    \includegraphics[width=\linewidth]{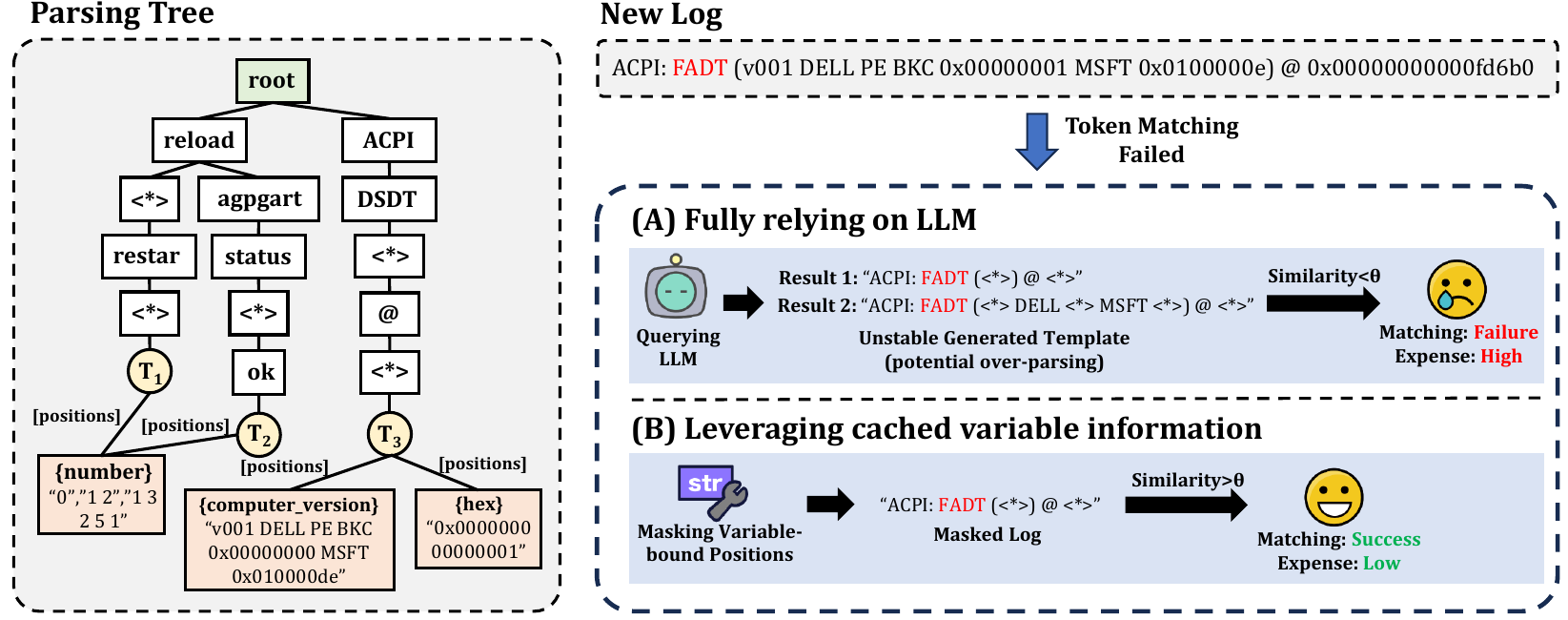}
    \caption{An illustrative case of variable omission in initial parsing, resulting in token matching failure for the log of the same template.}
    \label{fig:cache}
\end{figure}

For the logs that fail fuzzy matching, we perform a basic token-level match between the log and the parsing tree, following the methodology of prior works \cite{LILAC, logparser-llm}. However, when an unrecognized variable in the first parsing causes a new log token from the same template, token matching still fails. Existing methods directly query the LLM for a new template, which leads to redundant LLM invocations and, due to the instability of LLM outputs, also affects parsing accuracy to a certain degree. Figure \ref{fig:cache} illustrates a case where an unrecognized variable in the first parsing causes a new log token from the same template to fail in matching. In this case, existing methods directly query the LLM to generate the template. However, due to the inherent non-determinism of LLM outputs, this approach often leads to inconsistent parsing results for logs that belong to the same template.

As illustrated in part (A) of Figure \ref{fig:cache}, we find that the LLM produces unstable parsing results for similar logs, sometimes leading to over-parsing, which leads to low similarity scores and template matching failures. This not only undermines parsing accuracy but also incurs high costs due to repeated LLM invocations. In our approach, since the variable position information is stored in the cache, we can mask the bound variable positions in the corresponding template. This design significantly improves parsing accuracy while reducing redundant LLM invocations. Specifically, since token matching does not reach a leaf node, it terminates at one or more internal nodes. The templates within the subtrees of these stopping nodes are referred to as relevant templates, meaning the unparsed log partially matches these templates. Therefore, we traverse all relevant templates and perform similarity calculations with masking. Since the parsing cache stores the positional information of the variable parts in the template, for each template, we can mask all positions in the log token list that may correspond to variables, thereby achieving a fair match that is not influenced by known variables. The masked token list is then compared with each relevant template using Jaccard similarity. If the similarity exceeds a predefined threshold $\theta$, set to 0.8 in our implementation, it is considered a successful match. In this scenario, \name treats the inconsistent tokens as variables, creates a new variable unit with a label starting with 'unknown', and stores these inconsistent tokens within it.

If none of the relevant templates in the cache have a mask similarity with the new log that exceeds the threshold, it indicates that the parsing cache has failed to match the log. In this case, \name proceeds with adaptive variable-aware in-context learning to generate the log's template. Once the new template is generated, the prefix tree and the variable units in the cache are updated based on the template and its labeled variables. Specifically, new branches are created in the prefix tree starting from the stopping node, following the token list of the template. For each labeled variable, its corresponding variable unit is identified based on its label, and then the labeled variable information is added to the variable unit. If no matching variable unit is found, a new unit with the given label is created, and the variable's tokens are added to it. Finally, associations between the template and the variable units are established, storing the positional information of the variable parts.

\subsubsection{\textbf{Cache correction}}

In our experiments, we observed that large language models may misidentify the variable parts of logs due to the hallucination problem \cite{hallucation}. With the introduction of variable units, \name requires the LLM to assign a label to each variable, enabling us to perform preliminary corrections based on the label content. Specifically, whenever a new template is generated and the parsing cache needs to be updated, we validate the label of variable units. As illustrated in Table \ref{tab:hulla}, when the label is merely “label,” such as “exception in getting events $\rightarrow$ exception in getting \{label\}”; when the label is identical to the variable content and is a common word, such as “invalid uri in request $\rightarrow$ \{invalid\} uri in request”; and when the label is a punctuation mark, such as “i-cache parity error $\rightarrow$ i\{-\}cache parity error.” In these three cases, the identification of variables by LLMs is meaningless. Therefore, we corrected and converted them back into constants to ensure the accuracy of the parsing process.

\subsection{Adaptive Variable-aware ICL}

\begin{figure}[htp]
    \centering
    \includegraphics[width=\linewidth]{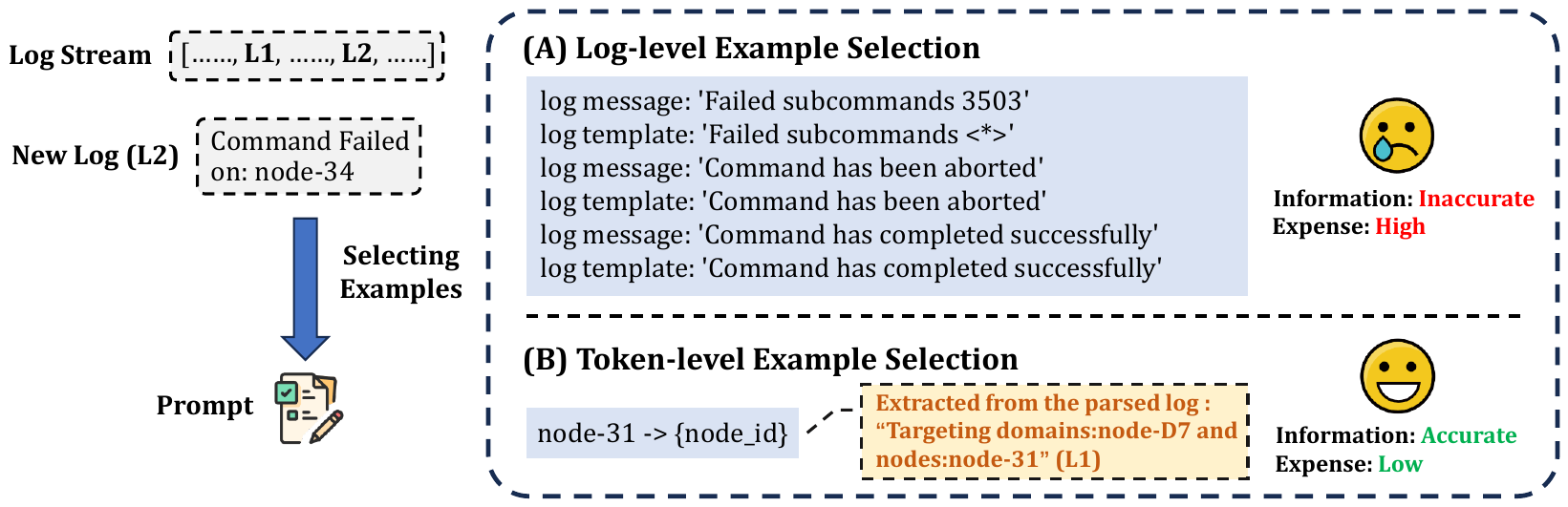}
    \caption{An illustrative case comparing example selection methods. In token-level selection, examples are extracted from historical logs of different templates (L1).}
    \label{fig:example}
\end{figure}

Logs that fail to match successfully in the parsing cache will be input into this step to obtain their parsing results. In the in-context learning step, existing methods select examples based on log-level similarity and provide log-template pair examples. However, since constants typically account for a large proportion of logs, the constant part often dominates the selection of examples. However, as LLMs can understand natural language, the repeated natural language in the examples contributes little to understanding the new logs to be parsed, resulting in token wastage and higher LLM invocation costs. To address this, we designed a token-level adaptive example selection and prompt format focused on variables. Figure \ref{fig:example} presents a case study comparing the example selection of constant-centric and variable-centric methods. As shown in part (B), the variable-centric approach not only significantly reduces token usage but also captures finer-grained information.

\subsubsection{\textbf{Token-level example selection}}

We first use the NLTK toolkit \cite{nltk} to extract the uncommon tokens in the new log. For each uncommon token, we calculate its Jaccard similarity with all tokens within the variable units stored in the parsing cache. The variable unit with the highest similarity is selected as an example, which is then presented to the LLM in the format "variable $\rightarrow$ \{label\}". Unlike existing methods that rely on a fixed number of examples, our approach adaptively determines the number of examples based on the number of uncommon tokens. It is important to note that, since labels starting with 'unknown' are meaningless, we exclude variable units with such labels from consideration in this step.

\subsubsection{\textbf{Prompt design}}

Following previous work \cite{DivLog, LILAC, logparser-llm}, we designed and employed the prompt format that consists of the following three parts:

\begin{enumerate}[leftmargin=*]
\item \textit{Instruction}. To ensure that the LLM is equipped with task-specific knowledge, we provide an instruction that briefly introduces the task, expected input and output formats.
\item \textit{Demonstration Examples}. The demonstration examples include variable examples and a format reference. The variable examples are selected using Adaptive Example Selection, guiding the LLM to identify variable parts in unseen logs. The format reference, chosen as the historical output with the highest Jaccard similarity to the new log, helps the LLM understand the context and adhere to the expected output format.
\item \textit{Queried Log}. This is the new log that needs to be parsed.
\end{enumerate}

The specific content of the prompt is shown in figure \ref{fig:prompt}. With the guidance of instructions and demonstration examples, the LLM can generate a more accurate parsed result for the queried log while maintaining the correct format.

\section{Experiment Setup}

\subsection{Datasets}

Our experiments are conducted using Loghub-2.0 \cite{loghub-2.0, loghub}, a widely recognized benchmark in the field of log parsing. Loghub-2.0 is a collection of large-scale datasets for log parsing provided by LogPAI \cite{LogPAI}, encompassing 14 log datasets sourced from diverse applications and systems. On average, each dataset in Loghub-2.0 contains approximately 3.69 million logs, all of which are annotated with ground truth log templates. 

\subsection{Baselines and Metrics}

Based on recent benchmark studies \cite{loghub-2.0, LogPAI-bench}, we compare \name with eight state-of-the-art baselines, including two syntax-based parsers, two semantic-based parsers, and four LLM-based parsers. In Appendix \ref{appd:baselines}, we briefly discuss all the baselines. Following recent studies \cite{LILAC, logbatcher, Lunar, logparser-llm}, we evaluate our methods using four metrics: Grouping Accuracy (GA) and Parsing Accuracy (PA) focus on message volume, while F1-score of Group Accuracy (FGA) and F1-score of Template Accuracy (FTA) provide fair template-level evaluation. The detailed definitions of evaluation metrics are listed in Appendix \ref{appd:metrics}.

\subsection{Implementation Details}

We perform our experiments on a laptop with an 8-core CPU and 16GB of memory. Apart from LibreLog, which uses Llama3-8b for efficiency, all other LLM-based baselines use gpt-3.5-turbo as the default LLM. To ensure fair evaluation and reproducibility, we use gpt-3.5-turbo via the OpenAI API with the temperature and seed both set to 0 for deterministic outputs. The source code and data can be accessed at https://github.com/mianmaner/VarParser. More specific implementation details are provided in Appendix \ref{appd:details}.


\section{Evaluation}

We evaluate \name by answering four research questions.

\begin{itemize}[leftmargin=*]
    \item RQ1: How does \name perform compared to baselines?
    \item RQ2: How do different modules contribute to \name?
    \item RQ3: How does \name perform on different LLMs?
    \item RQ4: How is the result integrity of \name?
\end{itemize}

\subsection{RQ1: Comparative Performance}

\subsubsection{\textbf{Accuracy}}

We evaluate \name and all baseline methods across all log datasets. The results are shown in Table \ref{tab:accuracy}, where the best and second-best results are in bold and underlined, respectively. Notably, the metric of AEL \cite{AEL} on the Spark dataset is marked with a '-' because, according to previous work \cite{LogPAI-bench, LILAC, logbatcher, Lunar}, it failed to complete the analysis process in a reasonable time (>240 hours). Specifically, compared to the highest performance of all baselines, \name outperforms by 3.9\% in GA, 8.5\% in PA, and 5.8\% in FTA on average. It is worth noting that although LogBatcher slightly outperforms \name on FGA, it performs far worse on other metrics. This suggests that while LogBatcher groups most templates correctly, the resulting groups contain very few logs and thus lack coverage. We further analyze the main reasons why \name does not surpass other baselines in some projects. Beyond the instability of LLM outputs, certain datasets contain constant values that closely resemble variables. For example, in MAC logs such as “en0: channel changed to 132,+1”, the token “en0” is actually a constant, but \name mistakenly treated it as an uncommon token and selected “0->{number}” as an exemplar, leading the LLM to misidentify it as a variable. Nevertheless, \name consistently achieves the most favorable accuracy across the majority of datasets. In particular, under the most rigorous and comprehensive metric, FTA, \name demonstrates superior performance over all baselines, highlighting its strong capability to accurately separate log templates from variables under strict evaluation criteria.

\begin{table*}[htbp]
\centering
\caption{Accuracy comparison between \name and baselines (\%). The best and second-best results are in bold and underlined.}
\label{tab:accuracy}
\resizebox{\textwidth}{!}{
    \begin{tabular}{ll||cccccccccccccc|c}
    \toprule[1pt]
    Method & Metric & Proxifier & Apache & OpenSSH & HDFS & OpenStack & HPC & Zookeeper & HealthApp & Hadoop & Spark & BGL & Linux & Mac & Thunderbird & Average \\
    \midrule
    \rowcolor{gray!20}
    \multicolumn{17}{c}{\textbf{Syntax-based Log Parsers}} \\
    \midrule
    \multirow{4}{*}{\textbf{AEL}}
     & GA  & 97.4 & \textbf{100.0} & 70.5 & \underline{99.9} & 74.3 & 74.8 & 99.6 & 72.5 & 82.3 & - & 91.5 & 91.6 & 79.7 & 78.6 & 85.6 \\
     & PA  & 67.7 & 72.7 & 36.4 & 62.1 & 2.9 & 74.1 & 84.2 & 31.1 & 53.5 & - & 40.6 & 8.2 & 24.5 & 16.3 & 44.2 \\
     & FGA & 66.7 & \textbf{100.0} & 68.9 & 76.4 & 68.2 & 20.1 & 78.8 & 0.8 & 11.7 & - & 58.7 & 80.6 & 79.3 & 11.6 & 55.5 \\
     & FTA & 41.7 & 51.7 & 33.3 & 56.2 & 16.5 & 13.6 & 46.5 & 0.3 & 5.8 & - & 16.5 & 21.7 & 20.5 & 3.5 & 25.2 \\
    \hline
    \multirow{4}{*}{\textbf{Drain}}
     & GA  & 69.2 & \textbf{100.0} & 70.7 & 99.9 & 75.2 & 79.3 & 99.4 & 86.2 & 92.1 & 88.8 & 91.9 & 68.6 & 76.1 & 83.1 & 84.3 \\
     & PA  & 68.8 & 72.7 & 58.6 & 62.1 & 2.9 & 72.1 & 84.3 & 31.2 & 54.1 & 39.4 & 40.7 & 11.1 & 35.7 & 21.6 & 46.8 \\
     & FGA & 20.6 & \textbf{100.0} & 87.2 & 93.5 & 0.7 & 30.9 & 90.4 & 1.0 & 78.5 & 86.1 & 62.4 & 77.8 & 22.9 & 23.7 & 55.4 \\
     & FTA & 17.6 & 51.7 & 48.7 & 60.9 & 0.2 & 15.2 & 61.4 & 0.4 & 38.4 & 41.2 & 19.3 & 25.9 & 6.9 & 7.1 & 28.2 \\
    \midrule
    \rowcolor{gray!20}
    \multicolumn{17}{c}{\textbf{Semantic-based Log Parsers}} \\
    \midrule
    \multirow{4}{*}{\textbf{Uniparser}}
     & GA  & 50.9 & 94.7 & 27.5 & \textbf{100.0} & \textbf{100.0} & 77.7 & 98.8 & 46.1 & 69.1 & 85.4 & 91.8 & 28.5 & 73.7 & 57.9 & 71.6 \\
     & PA  & 63.4 & 94.2 & 28.9 & 94.8 & 51.6 & 94.1 & \underline{98.8} & 81.7 & \underline{88.9} & 79.5 & 94.9 & 16.4 & 68.8 & 65.4 & 73.0 \\
     & FGA & 28.6 & 68.7 & 0.9 & \underline{96.8} & 96.9 & 66.0 & 66.1 & 74.5 & 62.8 & 2.0 & 62.4 & 45.1 & 69.9 & 68.2 & 57.8 \\
     & FTA & 45.7 & 26.9 & 0.5 & 58.1 & 28.9 & 35.1 & 51.0 & 46.2 & 47.6 & 1.2 & 21.9 & 23.2 & 28.3 & 29.0 & 31.7 \\
    \hline
    \multirow{4}{*}{\textbf{LogPPT}}
     & GA  & 98.9 & 78.6 & 27.7 & 72.1 & 53.4 & 78.2 & 96.7 & 99.8 & 48.3 & 47.6 & 24.5 & 20.5 & 54.4 & 56.4 & 61.2 \\
     & PA  & \textbf{100.0} & 94.8 & 65.4 & 94.3 & 40.6 & \textbf{99.7} & 84.5 & \textbf{99.7} & 66.6 & 95.2 & 93.8 & 16.8 & 39.0 & 40.1 & 73.6 \\
     & FGA & \underline{87.0} & 60.5 & 8.1 & 39.1 & 87.4 & 78.0 & 91.8 & 94.7 & 52.6 & 37.4 & 25.3 & 71.2 & 49.3 & 21.6 & 57.4 \\
     & FTA & \underline{95.7} & 36.8 & 10.5 & 31.2 & 73.8 & 76.8 & 80.9 & 82.2 & 43.4 & 29.9 & 26.1 & 42.8 & 27.4 & 11.7 & 47.8 \\
    \midrule
    \rowcolor{gray!20}
    \multicolumn{17}{c}{\textbf{LLM-based Log Parsers}} \\
    \midrule
    \multirow{4}{*}{\textbf{LibreLog}}
     & GA & 51.0 & \textbf{100.0} & \underline{86.8} & \textbf{100.0} & 81.1 & 84.4 & 99.3 & 86.2 & \textbf{96.3} & 85.9 & 90.2 & \textbf{91.2} & 81.4 & \textbf{87.0} & 87.2 \\
     & PA  & 89.7 & \underline{99.6} & 49.6 & \textbf{100.0} & 83.1 & 97.3 & 85.0 & \underline{97.4} & 87.1 & 88.9 & 92.9 & \textbf{90.2} & 65.4 & \textbf{69.4} & \underline{85.4} \\
     & FGA & 35.3 & \textbf{100.0} & 85.3 & 92.9 & 68.9 & 80.5 & \underline{97.4} & \underline{95.8} & \underline{94.0} & 83.2 & 80.3 & 74.1 & 80.4 & 84.3 & 82.3 \\
     & FTA & 40.0 & 75.9 & 47.2 & 79.5 & 55.9 & 60.9 & 86.3 & \textbf{87.7} & \underline{75.5} & 69.0 & 71.6 & 60.7 & 46.9 & 54.0 & 65.1 \\
    \hline
    \multirow{4}{*}{\textbf{LILAC}}
     & GA & 80.3 & \textbf{100.0} & 74.8 & \textbf{100.0} & \textbf{100.0} & 86.3 & 99.1 & \textbf{100.0} & 86.0 & \textbf{100.0} & 89.4 & 26.9 & \textbf{91.7} & 81.2 & 86.8 \\
     & PA  & \underline{91.6} & 99.5 & \underline{96.6} & \underline{99.9} & \textbf{100.0} & 92.9 & 82.6 & 55.1 & 83.3 & 99.8 & 94.4 & 30.0 & 69.7 & 57.7 & 82.4 \\
     & FGA & 64.0 & \textbf{100.0} & \underline{87.7} & \underline{96.8} & \textbf{100.0} & 84.9 & 93.0 & \textbf{98.1} & \textbf{91.8} & \textbf{90.8} & \textbf{90.8} & \textbf{91.4} & \textbf{92.4} & \textbf{88.7} & 90.7 \\
     & FTA & 88.0 & \underline{82.8} & \underline{79.5} & 90.3 & \textbf{95.8} & 67.6 & \underline{86.5} & 84.8 & \textbf{77.2} & \underline{74.7} & 76.4 & \textbf{72.3} & 60.3 & \textbf{57.9} & \underline{78.1} \\
    \hline
    \multirow{4}{*}{\textbf{LogBatcher}}
     & GA  & \textbf{100.0} & \underline{99.7} & \textbf{92.6} & \textbf{100.0} & 56.4 & \underline{86.4} & \underline{99.6} & 99.3 & \underline{93.7} & \underline{97.5} & 80.6 & 68.9 & 85.4 & 81.0 & \underline{88.7} \\
     & PA & \textbf{100.0} & 99.5 & 52.8 & 99.9 & 41.7 & 93.9 & 82.2 & 82.7 & 82.8 & \textbf{99.7} & 95.8 & 65.3 & 67.1 & 46.7 & 79.3 \\
     & FGA & \textbf{100.0} & \underline{94.9} & 87.3 & \textbf{100.0} & 96.9 & \textbf{89.7} & 94.1 & \textbf{98.1} & 90.4 & \textbf{90.8} & 90.4 & \underline{85.9} & \underline{89.5} & 82.1 & \textbf{92.2} \\
     & FTA & \textbf{100.0} & 81.4 & 59.2 & \underline{94.6} & 76.3 & \textbf{85.9} & 83.5 & \underline{86.5} & 73.9 & 73.9 & 55.3 & 65.8 & 55.3 & 46.9 & 74.2 \\
    \hline
    \multirow{4}{*}{\textbf{LUNAR}}
     & GA  & \underline{98.9} & \textbf{100.0} & 69.0 & 87.4 & 96.2 & 84.5 & 99.3 & \underline{99.7} & 92.4 & 85.1 & \underline{95.0} & 64.1 & 86.4 & 81.1 & \underline{88.7} \\
     & PA  & \textbf{100.0} & 82.4 & 71.9 & 94.0 & 87.4 & \underline{98.8} & 84.8 & 94.7 & 80.5 & 96.6 & \textbf{96.3} & 68.9 & 63.0 & 57.0 & 83.4 \\
     & FGA & \underline{87.0} & \textbf{100.0} & 68.8 & 85.7 & 25.6 & 72.1 & 88.5 & 95.0 & 78.5 & 81.4 & 86.6 & 80.4 & 75.9 & 82.8 & 79.8 \\
     & FTA & \underline{95.7} & 75.9 & 75.0 & 89.8 & 23.8 & 76.5 & 75.2 & 69.8 & 61.3 & 61.6 & 77.7 & 58.2 & 49.9 & 52.0 & 69.1 \\
    \hline
    \multirow{4}{*}{\textbf{\name}}
     & GA  & \textbf{100.0} & \textbf{100.0} & 74.8 & \textbf{100.0} & \textbf{100.0} & \textbf{87.0} & \textbf{100.0} & 86.5 & 90.3 & 95.5 & \textbf{96.9} & \underline{86.1} & \underline{91.0} & \underline{82.0} & \textbf{92.2} \\
     & PA  & \textbf{100.0} & \textbf{99.9} & \textbf{100.0} & \textbf{100.0} & \textbf{100.0} & \textbf{99.7} & \textbf{99.9} & 97.3 & \textbf{89.0} & \underline{98.4} & \underline{96.2} & \underline{81.2} & \textbf{71.7} & \underline{64.5} & \textbf{92.7} \\
     & FGA & \textbf{100.0} & \textbf{100.0} & \textbf{89.2} & \textbf{100.0} & \textbf{100.0} & \underline{85.9} & \textbf{100} & 90.5 & 88.8 & \underline{87.0} & \underline{90.6} & 82.5 & 85.6 & \underline{85.6} & \underline{91.8} \\
     & FTA & \textbf{100.0} & \textbf{89.7} & \textbf{91.9} & \textbf{97.8} & \textbf{95.8} & \underline{83.2} & \textbf{94.8} & 79.9 & 74.7 & \textbf{79.4} & \textbf{78.4} & \underline{70.9} & \textbf{60.7} & \textbf{57.9} & \textbf{82.6} \\
    \bottomrule[1pt]
    \end{tabular}}
\end{table*}

\subsubsection{\textbf{Efficiency}}

\begin{figure}[htp]
    \centering
    \includegraphics[width=\linewidth]{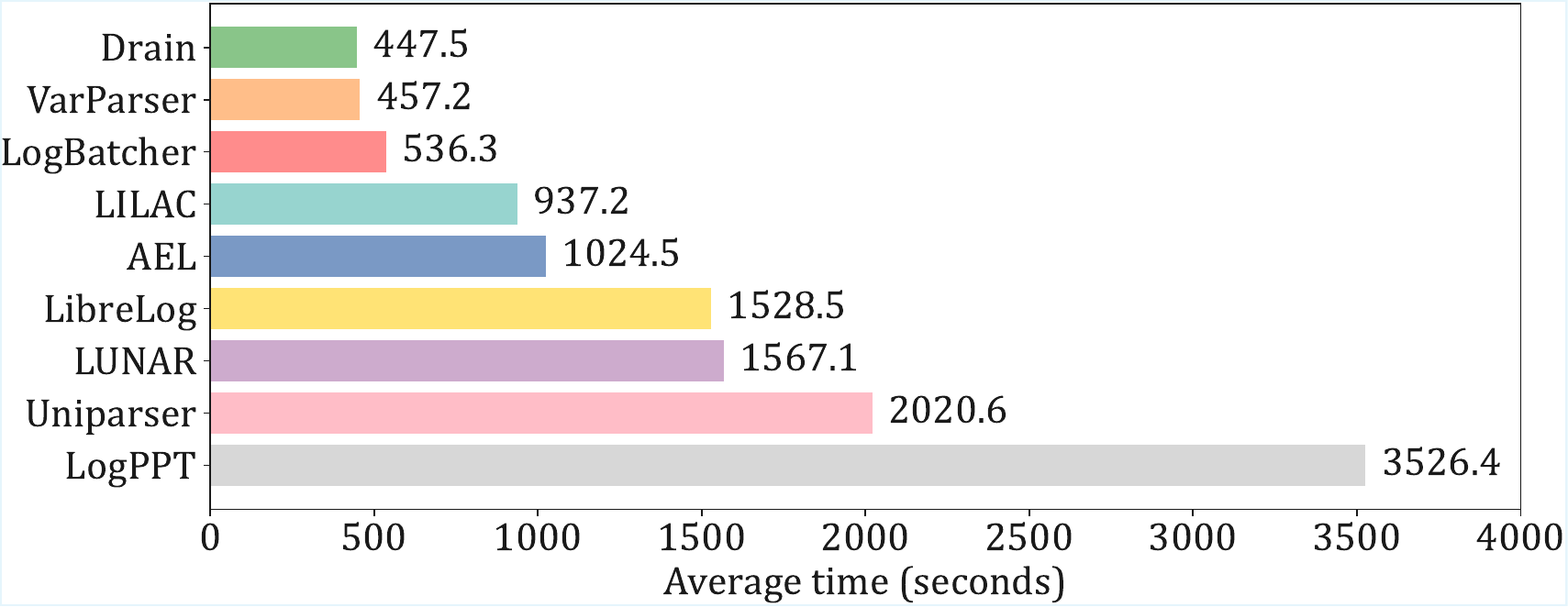}
    \caption{Average parsing time per dataset of \name and baselines (seconds).}
    \label{fig:effi}
\end{figure}

With the rapid growth in raw log volume, efficiency has always been a weak point for LLM-based parsers, limiting their practicality to some extent. Therefore, \name also focuses on improving efficiency. Since log sampling is a one-time process, we evaluated efficiency through experiments on both log sampling and log parsing. Due to fundamental methodological differences, some baselines \cite{logbatcher, Lunar, LibreLog} require splitting complete log datasets into chunks for processing. Therefore, we compared the average sampling time per dataset for \name and other LLM-based parsers under different sampling sizes, as shown in Table \ref{tab:sample_effi}. Specifically, in log sampling, compared with LILAC, our approach achieves an average speedup of $2.3\times$ across various sampling sizes. This improvement stems from \name reducing the sampling pool from the scale of templates to the scale of variables by grouping logs based on uncommon tokens. In addition, Figure \ref{fig:effi} shows the average log parsing time per dataset for \name and the baselines. Compared to LILAC, our method reduces time cost by 51.2\%, and compared to LogBatcher, it achieves a 14.8\% reduction, reaching efficiency levels very close to the syntax-based parser Drain. While both \name and most baselines employ prefix trees to accelerate template queries, \name further incorporates a fuzzy matching table to avoid redundant parsing of logs that differ only in common variables. Moreover, by reducing LLM invocation frequency and token usage, \name significantly improves log parsing efficiency.

\begin{table}[htp]
\centering
  \caption{Average sampling time of \name and baselines per dataset (seconds).}
  \label{tab:sample_effi}
  \begin{tabular}{lccccc}
      \toprule[1pt]
      & \textbf{8 logs} & \textbf{16 logs} & \textbf{32 logs} & \textbf{64 logs} & \textbf{128 logs} \\
      \midrule
      \textbf{LogPPT} & 284.1 & 629.3 & 1303.9 & 2779.6 & 5396.9 \\
      \textbf{LILAC} & 19.2 & 19.3 & 19.3 & 19.3 & 19.4 \\
      \textbf{\name} & 8.5 & 8.5 & 8.5 & 8.5 & 8.6 \\
      \bottomrule[1pt]
  \end{tabular}
\end{table}
\vspace{-3px}

\subsubsection{\textbf{LLM invocation costs}}

The expensive pricing of LLM APIs, particularly when processing large-scale log data, can significantly impact the economic feasibility and practicality of these methods. To address this, \name is optimized along two key dimensions: reducing the number of LLM invocations and lowering the average token consumption per invocation. The experimental results are presented in Table \ref{tab:llm_cost}, where token usage is reported by the OpenAI API and includes both input and output tokens.

\begin{table}[htp]
\setlength\tabcolsep{3pt}
\caption{Comparison of the number of Total tokens consumed and Average tokens consumed per invocation between \name and other LLM-based Log parsers. The best results are in bold.}
\label{tab:llm_cost}
    \scalebox{0.8}{
    \begin{tabular}{l|cc|cc|cc|cc}
        \toprule
        \multirow{2}{*}{} & \multicolumn{2}{c|}{\textbf{LUNAR}} & \multicolumn{2}{c|}{\textbf{LogBatcher}} & \multicolumn{2}{c|}{\textbf{LILAC}} & \multicolumn{2}{c}{\textbf{\name}} \\
        & Total & Avg. & Total & Avg. & Total & Avg. & Total & Avg. \\
        \hline
        Apache & 17639 & 551.2 & 3497 & \textbf{120.6} & 8499 & 294.0 & \textbf{1104} & 138.0 \\
        BGL & 247420 & 591.9 & 54583 & 157.3 & 90017 & 285.4 & \textbf{38810} & \textbf{141.1} \\
        Hadoop & 265317 & 647.1 & 45352 & 173.1 & 88941 & 388.1 & \textbf{40897} & \textbf{171.8} \\
        HDFS & 64252 & 676.3 & 15693 & 333.9 & 21314 & 453.5 & \textbf{4096} & \textbf{195.1} \\
        HealthAPP & 90187 & 530.5 & 16208 & \textbf{103.9} & 33390 & 315.0 & \textbf{11610} & 133.5 \\
        HPC & 85653 & 658.9 & 18052 & 214.9 & 14461 & 242.5 & \textbf{5757} & \textbf{122.5} \\
        Linux & 229202 & 520.9 & \textbf{40518} & \textbf{103.1} & 86763 & 265.2 & 52955 & 150.9 \\
        Mac & 1148074 & 699.2 & 153660 & 236.4 & 264647 & 349.6 & \textbf{151240} & \textbf{214.8} \\
        OpenSSH & 41990 & 567.4 & \textbf{5698} & \textbf{154.0} & 12535 & 358.1 & 8717 & 170.9 \\
        OpenStack & 200503 & 753.8 & 21609 & 441.0 & 21149 & 440.6 & \textbf{2984} & \textbf{175.5} \\
        Proxifier & 7058 & 588.2 & 4719 & 429.0 & 5511 & 393.5 & \textbf{655} & \textbf{218.3} \\
        Spark & 239019 & 627.3 & 42238 & 171.7 & 78058 & 351.6 & \textbf{34063} & \textbf{153.4} \\
        Thunderbird & 969604 & 572.0 & 201599 & \textbf{147.8} & 507329 & 349.4 & \textbf{191724} & 157.3 \\
        Zookeeper & 46855 & 557.8 & 10068 & \textbf{124.3} & 22919 & 248.8 & \textbf{6665} & 133.3 \\
        \hline
        Average & 260912.4 & 601.1 & 45250.0 & 207.9 & 89648.8 & 338.3 & \textbf{39377.2} & \textbf{162.6} \\
        \bottomrule
    \end{tabular}}
\end{table}

Specifically, compared to LILAC, our method reduces the total token consumption per dataset by an average of 56.1\%, making it a more economical and efficient solution for LLM-based log parsing. The latest study, LogBatcher, also lowers LLM invocation costs by processing all logs in batches and performing demonstration-free parsing through comparisons among similar logs. However, this requires collecting logs in advance, leading to system downtime during which newly generated logs cannot be parsed or immediately utilized by downstream tasks. In contrast, both LILAC and \name parse logs sequentially, making them more suitable for log stream scenarios. Remarkably, even against LogBatcher, \name achieves substantial savings, reducing token usage per invocation by 21.8\% and overall token consumption per dataset by 13.0\% on average. Such cost reductions are particularly significant for handling massive log volumes in real-world industrial environments.

\subsection{RQ2: Module Contribution Analysis}

To analyze the impact of different components on parsing accuracy, we conduct ablation studies on variable contribution sampling, mask matching, cache correction, and adaptive variable-aware ICL. For contribution sampling, we replace it with random sampling, and for variable-aware ICL, we remove the variable demonstration examples. The results of the ablation experiments are shown in Table \ref{tab:ablation}. The results indicate that each component positively contributes to the performance of the method. 

\begin{table}[htp]
\centering
  \caption{Ablation studies of \name (\%).}
  \label{tab:ablation}
  \renewcommand{\arraystretch}{0.9}
  \begin{tabular}{lcccc}
      \toprule[1pt]
      & \textbf{GA} & \textbf{FGA} & \textbf{PA} & \textbf{FTA} \\
      \midrule
      \textbf{\name} & 92.2 & 92.7 & 91.8 & 82.6 \\
      \textbf{w/o fuzzy matching} & 89.8 & 92.0 & 89.0 & 81.7 \\
      \textbf{w/o mask matching} & 88.1 & 88.9 & 83.3 & 80.5 \\
      \textbf{w/o cache correction} & 91.3 & 92.7 & 88.9 & 80.1 \\
      \textbf{w/o contribution sampling} & 87.9 & 81.4 & 80.5 & 70.0 \\
      \textbf{w/o variable-aware ICL} & 83.4 & 76.6 & 75.6 & 73.4 \\
      \bottomrule[1pt]
  \end{tabular}
\end{table}

When fuzzy matching is removed, \name's FTA drops by 1.1\% and FGA by 0.8\%, indicating that this strategy enables effective simple log matching, while also revealing that the LLM may still produce parsing errors on simple logs. Similarly, removing mask matching leads to a 2.1\% drop in FTA and a 2.8\% drop in FGA, indicating that mask matching helps the cache achieve more accurate matches and reduces the impact of LLM's unstable output. Furthermore, eliminating variable examples results in a 10.7\% decrease in FTA and a 16.3\% decrease in FGA, suggesting that LLMs' ability to handle diverse log data remains limited without domain-specific knowledge. Notably, replacing variable contribution sampling with random sampling reduces FTA by 14.8\% and FGA by 11.0\%, underscoring the importance of high-quality candidate samples and demonstrations in driving LLM performance. Finally, removing cache correction also causes a noticeable accuracy drop, highlighting the necessity of addressing the LLM hallucination problem.

\subsection{RQ3: Performance Across LLMs}
\label{appd:different llms}

In our experiments, we utilize gpt-3.5-turbo as the default LLM to evaluate the performance of \name. To further investigate the adaptability of \name across different LLMs, we also conduct experiments with two additional commonly used LLMs: llama3-70b \cite{llama3} and qwen-plus \cite{qwen-plus}. The experimental results are presented in Table \ref{tab:different_llms}. The results demonstrate that \name consistently outperforms all baselines across all four metrics (GA, FGA, PA, and FTA) for each of the tested LLMs. Notably, the performance of \name remains relatively stable across different models. Specifically, the FGA metric ranges from 90.7\% to 92.7\%, while the FTA metric shows values between 78.6\% and 82.6\%. The results highlight the adaptability of \name, suggesting that it can be effectively integrated with various LLMs to achieve high accuracy.

\begin{table}[htp]
\centering
  \caption{Accuracy of \name across different LLMs (\%).}
  \label{tab:different_llms}
  \renewcommand{\arraystretch}{0.9}
  \begin{tabular}{lcccc}
      \toprule[1pt]
      & \textbf{GA} & \textbf{FGA} & \textbf{PA} & \textbf{FTA} \\
      \midrule
      \textbf{gpt-3.5-turbo} & 92.2 & 92.7 & 91.8 & 82.6 \\
      \textbf{llama3-70b} & 89.3 & 90.7 & 85.1 & 78.6 \\
      \textbf{qwen-plus} & 91.2 & 92.7 & 88.9 & 80.1 \\
      \bottomrule[1pt]
  \end{tabular}
\end{table}

\subsection{RQ4: Result Integrity}

By introducing and caching variable units, \name also provides variable-related information across the entire log set. Figure \ref{fig:variable} presents the different types of variables extracted from the variable units after parsing the BGL dataset, along with their frequencies. We observe that variables such as \textit{file\_path} and \textit{file\_name} appear most frequently, suggesting that file operations are commonly involved during system operation. Hardware-related variables, including \textit{node\_id}, \textit{chip\_status}, and \textit{slot\_id}, also occur extensively, reflecting their importance in monitoring system states. In contrast, variables such as \textit{exception} or \textit{error\_message} are relatively rare, indicating that the system generally operates in a relatively stable state.

\section{Conclusion and Future work}

In this paper, we introduce \name, the first variable-centric LLM-based log parsing approach that emphasizes the variable parts of logs to reveal their contribution to parsing. By integrating variable contribution sampling, the variable-centric parsing cache, and adaptive variable-aware ICL, \name improves both accuracy and efficiency, while preserving richer variable-related information to enhance system visibility for engineers. In the future, We aim to follow the idea of our variable-centric strategy and consider combining the abilities of LLMs and small models to further improve the performance of our approach.

\section*{Acknowledgement}
This work was supported by the National Natural Science Foundation of China(Grant No. 62502015).

\bibliographystyle{ACM-Reference-Format}
\balance  
\bibliography{sample-base}

@article{nltk,
  author = {Edward Loper and
                  Steven Bird},
  title = {{NLTK:} The Natural Language Toolkit},
  journal = {CoRR},
  volume = {cs.CL/0205028},
  year = {2002},
  url = {https://arxiv.org/abs/cs/0205028},
  bibsource = {dblp computer science bibliography, https://dblp.org}
}

@INPROCEEDINGS {loghub,
author = { Zhu, Jieming and He, Shilin and He, Pinjia and Liu, Jinyang and Lyu, Michael R. },
booktitle = { 2023 IEEE 34th International Symposium on Software Reliability Engineering (ISSRE) },
title = {{ Loghub: A Large Collection of System Log Datasets for AI-driven Log Analytics }},
year = {2023},
volume = {},
ISSN = {},
pages = {355-366},
doi = {10.1109/ISSRE59848.2023.00071},
url = {https://doi.ieeecomputersociety.org/10.1109/ISSRE59848.2023.00071},
publisher = {IEEE Computer Society},
address = {Los Alamitos, CA, USA},
month =Oct}

@inproceedings{loghub-2.0,
author = {Jiang, Zhihan and Liu, Jinyang and Huang, Junjie and Li, Yichen and Huo, Yintong and Gu, Jiazhen and Chen, Zhuangbin and Zhu, Jieming and Lyu, Michael R.},
title = {A Large-Scale Evaluation for Log Parsing Techniques: How Far Are We?},
year = {2024},
isbn = {9798400706127},
publisher = {Association for Computing Machinery},
address = {New York, NY, USA},
url = {https://doi.org/10.1145/3650212.3652123},
doi = {10.1145/3650212.3652123},
booktitle = {Proceedings of the 33rd ACM SIGSOFT International Symposium on Software Testing and Analysis},
pages = {223–234},
numpages = {12},
location = {Vienna, Austria},
series = {ISSTA 2024}
}

@inproceedings{LogPAI-bench,
author = {Khan, Zanis Ali and Shin, Donghwan and Bianculli, Domenico and Briand, Lionel},
title = {Guidelines for assessing the accuracy of log message template identification techniques},
year = {2022},
isbn = {9781450392211},
publisher = {Association for Computing Machinery},
address = {New York, NY, USA},
url = {https://doi.org/10.1145/3510003.3510101},
doi = {10.1145/3510003.3510101},
booktitle = {Proceedings of the 44th International Conference on Software Engineering},
pages = {1095–1106},
numpages = {12},
location = {Pittsburgh, Pennsylvania},
series = {ICSE '22}
}

@inproceedings{LogPAI,
author = {Zhu, Jieming and He, Shilin and Liu, Jinyang and He, Pinjia and Xie, Qi and Zheng, Zibin and Lyu, Michael R.},
title = {Tools and benchmarks for automated log parsing},
year = {2019},
publisher = {IEEE Press},
url = {https://doi.org/10.1109/ICSE-SEIP.2019.00021},
doi = {10.1109/ICSE-SEIP.2019.00021},
booktitle = {Proceedings of the 41st International Conference on Software Engineering: Software Engineering in Practice},
pages = {121–130},
numpages = {10},
location = {Montreal, Quebec, Canada},
series = {ICSE-SEIP '19}
}

@INPROCEEDINGS{AEL,
  author={Jiang, Zhen Ming and Hassan, Ahmed E. and Flora, Parminder and Hamann, Gilbert},
  booktitle={2008 The Eighth International Conference on Quality Software}, 
  title={Abstracting Execution Logs to Execution Events for Enterprise Applications (Short Paper)}, 
  year={2008},
  volume={},
  number={},
  pages={181-186},
  doi={10.1109/QSIC.2008.50}}

@INPROCEEDINGS{Drain,
  author={He, Pinjia and Zhu, Jieming and Zheng, Zibin and Lyu, Michael R.},
  booktitle={2017 IEEE International Conference on Web Services (ICWS)}, 
  title={Drain: An Online Log Parsing Approach with Fixed Depth Tree}, 
  year={2017},
  volume={},
  number={},
  pages={33-40},
  doi={10.1109/ICWS.2017.13}}

@inproceedings{Uniparser,
author = {Liu, Yudong and Zhang, Xu and He, Shilin and Zhang, Hongyu and Li, Liqun and Kang, Yu and Xu, Yong and Ma, Minghua and Lin, Qingwei and Dang, Yingnong and Rajmohan, Saravan and Zhang, Dongmei},
title = {UniParser: A Unified Log Parser for Heterogeneous Log Data},
year = {2022},
isbn = {9781450390965},
publisher = {Association for Computing Machinery},
address = {New York, NY, USA},
url = {https://doi.org/10.1145/3485447.3511993},
doi = {10.1145/3485447.3511993},
booktitle = {Proceedings of the ACM Web Conference 2022},
pages = {1893–1901},
numpages = {9},
location = {Virtual Event, Lyon, France},
series = {WWW '22}
}

@inproceedings{LogPPT,
author = {Le, Van-Hoang and Zhang, Hongyu},
title = {Log Parsing with Prompt-Based Few-Shot Learning},
year = {2023},
isbn = {9781665457019},
publisher = {IEEE Press},
url = {https://doi.org/10.1109/ICSE48619.2023.00204},
doi = {10.1109/ICSE48619.2023.00204},
booktitle = {Proceedings of the 45th International Conference on Software Engineering},
pages = {2438–2449},
numpages = {12},
location = {Melbourne, Victoria, Australia},
series = {ICSE '23}
}

@article{LILAC,
author = {Jiang, Zhihan and Liu, Jinyang and Chen, Zhuangbin and Li, Yichen and Huang, Junjie and Huo, Yintong and He, Pinjia and Gu, Jiazhen and Lyu, Michael R.},
title = {LILAC: Log Parsing using LLMs with Adaptive Parsing Cache},
year = {2024},
issue_date = {July 2024},
publisher = {Association for Computing Machinery},
address = {New York, NY, USA},
volume = {1},
number = {FSE},
url = {https://doi.org/10.1145/3643733},
doi = {10.1145/3643733},
journal = {Proc. ACM Softw. Eng.},
month = jul,
articleno = {7},
numpages = {24}
}

@inproceedings{DivLog,
author = {Xu, Junjielong and Yang, Ruichun and Huo, Yintong and Zhang, Chengyu and He, Pinjia},
title = {DivLog: Log Parsing with Prompt Enhanced In-Context Learning},
year = {2024},
isbn = {9798400702174},
publisher = {Association for Computing Machinery},
address = {New York, NY, USA},
url = {https://doi.org/10.1145/3597503.3639155},
doi = {10.1145/3597503.3639155},
booktitle = {Proceedings of the IEEE/ACM 46th International Conference on Software Engineering},
articleno = {199},
numpages = {12},
location = {Lisbon, Portugal},
series = {ICSE '24}
}

@inproceedings{DeepLog,
author = {Du, Min and Li, Feifei and Zheng, Guineng and Srikumar, Vivek},
title = {DeepLog: Anomaly Detection and Diagnosis from System Logs through Deep Learning},
year = {2017},
isbn = {9781450349468},
publisher = {Association for Computing Machinery},
address = {New York, NY, USA},
url = {https://doi.org/10.1145/3133956.3134015},
doi = {10.1145/3133956.3134015},
booktitle = {Proceedings of the 2017 ACM SIGSAC Conference on Computer and Communications Security},
pages = {1285–1298},
numpages = {14},
location = {Dallas, Texas, USA},
series = {CCS '17}
}

@INPROCEEDINGS{Midlog,
  author={He, Minghua and Jia, Tong and Duan, Chiming and Cai, Huaqian and Li, Ying and Huang, Gang},
  booktitle={2025 IEEE/ACM 47th International Conference on Software Engineering (ICSE)}, 
  title={Weakly-Supervised Log-Based Anomaly Detection with Inexact Labels via Multi-Instance Learning}, 
  year={2025},
  volume={},
  number={},
  pages={2918-2930},
  doi={10.1109/ICSE55347.2025.00189}}

@inproceedings{MetaLog,
author = {Zhang, Chenyangguang and Jia, Tong and Shen, Guopeng and Zhu, Pinyan and Li, Ying},
title = {MetaLog: Generalizable Cross-System Anomaly Detection from Logs with Meta-Learning},
year = {2024},
isbn = {9798400702174},
publisher = {Association for Computing Machinery},
address = {New York, NY, USA},
url = {https://doi.org/10.1145/3597503.3639205},
doi = {10.1145/3597503.3639205},
booktitle = {Proceedings of the IEEE/ACM 46th International Conference on Software Engineering},
articleno = {154},
numpages = {12},
location = {Lisbon, Portugal},
series = {ICSE '24}
}

@inproceedings{Medicine,
author = {Tao, Lei and Zhang, Shenglin and Jia, Zedong and Sun, Jinrui and Ma, Minghua and Li, Zhengdan and Yongqian, Sun and Yang, Canqun and Zhang, Yuzhi and Pei, Dan},
year = {2024},
booktitle = {Proceedings of the IEEE/ACM International Conference on Automated Software Engineering},
month = {10},
pages = {1107-1119},
title = {Giving Every Modality a Voice in Microservice Failure Diagnosis via Multimodal Adaptive Optimization},
doi = {10.1145/3691620.3695489}
}

@INPROCEEDINGS{cloud19,
  author={Yuan, Yue and Shi, Wenchang and Liang, Bin and Qin, Bo},
  booktitle={2019 IEEE 12th International Conference on Cloud Computing (CLOUD)}, 
  title={An Approach to Cloud Execution Failure Diagnosis Based on Exception Logs in OpenStack}, 
  year={2019},
  volume={},
  number={},
  pages={124-131},
  doi={10.1109/CLOUD.2019.00031}}

@INPROCEEDINGS{logcluster,
  author={Lin, Qingwei and Zhang, Hongyu and Lou, Jian-Guang and Zhang, Yu and Chen, Xuewei},
  booktitle={2016 IEEE/ACM 38th International Conference on Software Engineering Companion (ICSE-C)}, 
  title={Log Clustering Based Problem Identification for Online Service Systems}, 
  year={2016},
  volume={},
  number={},
  pages={102-111},
  doi={}}

@INPROCEEDINGS{logflatter,
  author={Amar, Anunay and Rigby, Peter C.},
  booktitle={2019 IEEE/ACM 41st International Conference on Software Engineering (ICSE)}, 
  title={Mining Historical Test Logs to Predict Bugs and Localize Faults in the Test Logs}, 
  year={2019},
  volume={},
  number={},
  pages={140-151},
  doi={10.1109/ICSE.2019.00031}}

@article{LogRule,
author = {Notaro, Paolo and Haeri, Soroush and Cardoso, Jorge and Gerndt, Michael},
title = {LogRule: Efficient Structured Log Mining for Root Cause Analysis},
year = {2023},
issue_date = {Dec. 2023},
publisher = {IEEE Press},
volume = {20},
number = {4},
issn = {1932-4537},
url = {https://doi.org/10.1109/TNSM.2023.3282270},
doi = {10.1109/TNSM.2023.3282270},
month = jun,
pages = {4231–4243},
numpages = {13}
}

@INPROCEEDINGS{logcluster-row,
  author={Vaarandi, Risto and Pihelgas, Mauno},
  booktitle={2015 11th International Conference on Network and Service Management (CNSM)}, 
  title={LogCluster - A data clustering and pattern mining algorithm for event logs}, 
  year={2015},
  volume={},
  number={},
  pages={1-7},
  doi={10.1109/CNSM.2015.7367331}}

@inproceedings{Hue,
author = {Xu, Junjielong and Fu, Qiuai and Zhu, Zhouruixing and Cheng, Yutong and Li, Zhijing and Ma, Yuchi and He, Pinjia},
title = {Hue: A User-Adaptive Parser for Hybrid Logs},
year = {2023},
isbn = {9798400703270},
publisher = {Association for Computing Machinery},
address = {New York, NY, USA},
url = {https://doi.org/10.1145/3611643.3616260},
doi = {10.1145/3611643.3616260},
booktitle = {Proceedings of the 31st ACM Joint European Software Engineering Conference and Symposium on the Foundations of Software Engineering},
pages = {413–424},
numpages = {12},
location = {San Francisco, CA, USA},
series = {ESEC/FSE 2023}
}

@ARTICLE{Brain,
  author={Yu, Siyu and He, Pinjia and Chen, Ningjiang and Wu, Yifan},
  journal={IEEE Transactions on Services Computing}, 
  title={Brain: Log Parsing With Bidirectional Parallel Tree}, 
  year={2023},
  volume={16},
  number={5},
  pages={3224-3237},
  doi={10.1109/TSC.2023.3270566}}

@INPROCEEDINGS {LLMParser,
author = { Ma, Zeyang and Chen, An Ran and Kim, Dong Jae and Chen, Tse-Hsun Peter and Wang, Shaowei },
booktitle = { 2024 IEEE/ACM 46th International Conference on Software Engineering (ICSE) },
title = {{ LLMParser: An Exploratory Study on Using Large Language Models for Log Parsing }},
year = {2024},
volume = {},
ISSN = {},
pages = {1209-1221},
doi = {10.1145/3597503.3639150},
url = {https://doi.ieeecomputersociety.org/10.1145/3597503.3639150},
publisher = {IEEE Computer Society},
address = {Los Alamitos, CA, USA},
month =apr}

@inproceedings{HumanLog,
author = {Jia, Tong and Li, Ying and Yang, Yong and Huang, Gang and Wu, Zhonghai},
title = {Augmenting Log-based Anomaly Detection Models to Reduce False Anomalies with Human Feedback},
year = {2022},
isbn = {9781450393850},
publisher = {Association for Computing Machinery},
address = {New York, NY, USA},
url = {https://doi.org/10.1145/3534678.3539106},
doi = {10.1145/3534678.3539106},
booktitle = {Proceedings of the 28th ACM SIGKDD Conference on Knowledge Discovery and Data Mining},
pages = {3081–3089},
numpages = {9},
location = {Washington DC, USA},
series = {KDD '22}
}

@inproceedings{codet5,
    title = "{C}ode{T}5: Identifier-aware Unified Pre-trained Encoder-Decoder Models for Code Understanding and Generation",
    author = "Wang, Yue  and
      Wang, Weishi  and
      Joty, Shafiq  and
      Hoi, Steven C.H.",
    editor = "Moens, Marie-Francine  and
      Huang, Xuanjing  and
      Specia, Lucia  and
      Yih, Scott Wen-tau",
    booktitle = "Proceedings of the 2021 Conference on Empirical Methods in Natural Language Processing",
    month = nov,
    year = "2021",
    address = "Online and Punta Cana, Dominican Republic",
    publisher = "Association for Computational Linguistics",
    url = "https://aclanthology.org/2021.emnlp-main.685",
    doi = "10.18653/v1/2021.emnlp-main.685",
    pages = "8696--8708",
}

@misc{ICL-survey1,
      title={A Survey on In-context Learning}, 
      author={Qingxiu Dong and Lei Li and Damai Dai and Ce Zheng and Jingyuan Ma and Rui Li and Heming Xia and Jingjing Xu and Zhiyong Wu and Tianyu Liu and Baobao Chang and Xu Sun and Lei Li and Zhifang Sui},
      year={2024},
      eprint={2301.00234},
      archivePrefix={arXiv},
      primaryClass={cs.CL},
      url={https://arxiv.org/abs/2301.00234}, 
}

@article{ICL-survey2,
author = {Liu, Pengfei and Yuan, Weizhe and Fu, Jinlan and Jiang, Zhengbao and Hayashi, Hiroaki and Neubig, Graham},
year = {2022},
month = {09},
pages = {},
title = {Pre-train, Prompt, and Predict: A Systematic Survey of Prompting Methods in Natural Language Processing},
volume = {55},
journal = {ACM Computing Surveys},
doi = {10.1145/3560815}
}

@article{hilogx,
author = {Jia, Tong and Li, Ying and Yang, Yong and Huang, Gang},
year = {2024},
month = {03},
pages = {1-18},
title = {Hilogx: noise-aware log-based anomaly detection with human feedback},
volume = {33},
journal = {The VLDB Journal},
doi = {10.1007/s00778-024-00843-2}
}

@INPROCEEDINGS{Toomanycooks,
  author={Zhang, Shenglin and Feng, Xiaoyu and Wang, Runzhou and Ma, Minghua and Gu, Wenwei and Sun, Yongqian and Jia, Zedong and Sun, Jinrui and Pei, Dan},
  booktitle={2025 IEEE 36th International Symposium on Software Reliability Engineering (ISSRE)}, 
  title={Too Many Cooks: Assessing the Need for Multi-Source Data in Microservice Failure Diagnosis}, 
  year={2025},
  volume={},
  number={},
  pages={1-12},
  doi={10.1109/ISSRE66568.2025.00014}}

@inproceedings{CloudRCA,
author = {Zhang, Yingying and Guan, Zhengxiong and Qian, Huajie and Xu, Leili and Liu, Hengbo and Wen, Qingsong and Sun, Liang and Jiang, Junwei and Fan, Lunting and Ke, Min},
title = {CloudRCA: A Root Cause Analysis Framework for Cloud Computing Platforms},
year = {2021},
isbn = {9781450384469},
publisher = {Association for Computing Machinery},
address = {New York, NY, USA},
url = {https://doi.org/10.1145/3459637.3481903},
doi = {10.1145/3459637.3481903},
booktitle = {Proceedings of the 30th ACM International Conference on Information \& Knowledge Management},
pages = {4373–4382},
numpages = {10},
location = {Virtual Event, Queensland, Australia},
series = {CIKM '21}
}

@INPROCEEDINGS{LogSed,
  author={Jia, Tong and Yang, Lin and Chen, Pengfei and Li, Ying and Meng, Fanjing and Xu, Jingmin},
  booktitle={2017 IEEE 10th International Conference on Cloud Computing (CLOUD)}, 
  title={LogSed: Anomaly Diagnosis through Mining Time-Weighted Control Flow Graph in Logs}, 
  year={2017},
  volume={},
  number={},
  pages={447-455},
  doi={10.1109/CLOUD.2017.64}}

@INPROCEEDINGS{LogFlash,
  author={Jia, Tong and Wu, Yifan and Hou, Chuanjia and Li, Ying},
  booktitle={2021 IEEE 32nd International Symposium on Software Reliability Engineering (ISSRE)}, 
  title={LogFlash: Real-time Streaming Anomaly Detection and Diagnosis from System Logs for Large-scale Software Systems}, 
  year={2021},
  volume={},
  number={},
  pages={80-90},
  doi={10.1109/ISSRE52982.2021.00021}}

@article{ICL-medi-1,
        title={ The Scope of In-Context Learning for the Extraction of Medical Temporal Constraints },
        author={ Parker Seegmiller and Joseph Gatto and Madhusudan Basak and Diane J. Cook and Hassan Ghasemzadeh and John A. Stankovic and Sarah Masud Preum },
        year={ 2023 },
        doi={ 10.1109/ichi57859.2023.00107 }}

@misc{ICL-sensi-1,
      title={On the Relation between Sensitivity and Accuracy in In-context Learning}, 
      author={Yanda Chen and Chen Zhao and Zhou Yu and Kathleen McKeown and He He},
      year={2024},
      eprint={2209.07661},
      archivePrefix={arXiv},
      primaryClass={cs.CL},
      url={https://arxiv.org/abs/2209.07661}, 
}

@misc{VALB,
      title={Did We Miss Something Important? Studying and Exploring Variable-Aware Log Abstraction}, 
      author={Zhenhao Li and Chuan Luo and Tse-Hsun Chen and Weiyi Shang and Shilin He and Qingwei Lin and Dongmei Zhang},
      year={2023},
      eprint={2304.11391},
      archivePrefix={arXiv},
      primaryClass={cs.SE},
      url={https://arxiv.org/abs/2304.11391}, 
}

@misc{qwen-plus,
      title={Qwen Technical Report}, 
      author={Jinze Bai and Shuai Bai and Yunfei Chu and Zeyu Cui and Kai Dang and Xiaodong Deng and Yang Fan and Wenbin Ge and Yu Han and Fei Huang and Binyuan Hui and Luo Ji and Mei Li and Junyang Lin and Runji Lin and Dayiheng Liu and Gao Liu and Chengqiang Lu and Keming Lu and Jianxin Ma and Rui Men and Xingzhang Ren and Xuancheng Ren and Chuanqi Tan and Sinan Tan and Jianhong Tu and Peng Wang and Shijie Wang and Wei Wang and Shengguang Wu and Benfeng Xu and Jin Xu and An Yang and Hao Yang and Jian Yang and Shusheng Yang and Yang Yao and Bowen Yu and Hongyi Yuan and Zheng Yuan and Jianwei Zhang and Xingxuan Zhang and Yichang Zhang and Zhenru Zhang and Chang Zhou and Jingren Zhou and Xiaohuan Zhou and Tianhang Zhu},
      year={2023},
      eprint={2309.16609},
      archivePrefix={arXiv},
      primaryClass={cs.CL},
      url={https://arxiv.org/abs/2309.16609}, 
}

@article{bi-lstm,
author = {Graves, Alex and Schmidhuber, J\"{u}rgen},
title = {2005 Special Issue: Framewise phoneme classification with bidirectional LSTM and other neural network architectures},
year = {2005},
issue_date = {June 2005},
publisher = {Elsevier Science Ltd.},
address = {GBR},
volume = {18},
number = {5–6},
issn = {0893-6080},
url = {https://doi.org/10.1016/j.neunet.2005.06.042},
doi = {10.1016/j.neunet.2005.06.042},
journal = {Neural Netw.},
month = jun,
pages = {602–610},
numpages = {9}
}

@misc{roberta,
      title={RoBERTa: A Robustly Optimized BERT Pretraining Approach}, 
      author={Yinhan Liu and Myle Ott and Naman Goyal and Jingfei Du and Mandar Joshi and Danqi Chen and Omer Levy and Mike Lewis and Luke Zettlemoyer and Veselin Stoyanov},
      year={2019},
      eprint={1907.11692},
      archivePrefix={arXiv},
      primaryClass={cs.CL},
      url={https://arxiv.org/abs/1907.11692}, 
}

@article{hallucation,
author = {Huang, Lei and Yu, Weijiang and Ma, Weitao and Zhong, Weihong and Feng, Zhangyin and Wang, Haotian and Chen, Qianglong and Peng, Weihua and Feng, Xiaocheng and Qin, Bing and Liu, Ting},
title = {A Survey on Hallucination in Large Language Models: Principles, Taxonomy, Challenges, and Open Questions},
year = {2024},
publisher = {Association for Computing Machinery},
address = {New York, NY, USA},
issn = {1046-8188},
url = {https://doi.org/10.1145/3703155},
doi = {10.1145/3703155},
note = {Just Accepted},
journal = {ACM Trans. Inf. Syst.},
month = nov
}

@inproceedings{logparser-llm,
author = {Zhong, Aoxiao and Mo, Dengyao and Liu, Guiyang and Liu, Jinbu and Lu, Qingda and Zhou, Qi and Wu, Jiesheng and Li, Quanzheng and Wen, Qingsong},
title = {LogParser-LLM: Advancing Efficient Log Parsing with Large Language Models},
year = {2024},
isbn = {9798400704901},
publisher = {Association for Computing Machinery},
address = {New York, NY, USA},
url = {https://doi.org/10.1145/3637528.3671810},
doi = {10.1145/3637528.3671810},
booktitle = {Proceedings of the 30th ACM SIGKDD Conference on Knowledge Discovery and Data Mining},
pages = {4559–4570},
numpages = {12},
location = {Barcelona, Spain},
series = {KDD '24}
}

@misc{logbatcher,
      title={Stronger, Cheaper and Demonstration-Free Log Parsing with LLMs}, 
      author={Yi Xiao and Van-Hoang Le and Hongyu Zhang},
      year={2024},
      eprint={2406.06156},
      archivePrefix={arXiv},
      primaryClass={cs.SE},
      url={https://arxiv.org/abs/2406.06156}, 
}

@misc{Lunar,
      title={LUNAR: Unsupervised LLM-based Log Parsing}, 
      author={Junjie Huang and Zhihan Jiang and Zhuangbin Chen and Michael R. Lyu},
      year={2024},
      eprint={2406.07174},
      archivePrefix={arXiv},
      primaryClass={cs.SE},
      url={https://arxiv.org/abs/2406.07174}, 
}

@misc{LibreLog,
      title={LibreLog: Accurate and Efficient Unsupervised Log Parsing Using Open-Source Large Language Models}, 
      author={Zeyang Ma and Dong Jae Kim and Tse-Hsun Chen},
      year={2024},
      eprint={2408.01585},
      archivePrefix={arXiv},
      primaryClass={cs.SE},
      url={https://arxiv.org/abs/2408.01585}, 
}

@misc{bi-lstm-crf,
      title={Bidirectional LSTM-CRF Models for Sequence Tagging}, 
      author={Zhiheng Huang and Wei Xu and Kai Yu},
      year={2015},
      eprint={1508.01991},
      archivePrefix={arXiv},
      primaryClass={cs.CL},
      url={https://arxiv.org/abs/1508.01991}, 
}

@misc{llama3,
      title={The Llama 3 Herd of Models}, 
      author={Aaron Grattafiori and Abhimanyu Dubey etc.},
      year={2024},
      eprint={2407.21783},
      archivePrefix={arXiv},
      primaryClass={cs.AI},
      url={https://arxiv.org/abs/2407.21783}, 
}

@inproceedings{onion,
author = {Zhang, Xu and Xu, Yong and Qin, Si and He, Shilin and Qiao, Bo and Li, Ze and Zhang, Hongyu and Li, Xukun and Dang, Yingnong and Lin, Qingwei and Chintalapati, Murali and Rajmohan, Saravanakumar and Zhang, Dongmei},
title = {Onion: identifying incident-indicating logs for cloud systems},
year = {2021},
isbn = {9781450385626},
publisher = {Association for Computing Machinery},
address = {New York, NY, USA},
url = {https://doi.org/10.1145/3468264.3473919},
doi = {10.1145/3468264.3473919},
booktitle = {Proceedings of the 29th ACM Joint Meeting on European Software Engineering Conference and Symposium on the Foundations of Software Engineering},
pages = {1253–1263},
numpages = {11},
location = {Athens, Greece},
series = {ESEC/FSE 2021}
}

@inbook{vista,
author = {Sun, Jinrui and Jia, Tong and He, Minghua and Wu, Yihan and Li, Ying and Huang, Gang},
title = {Exploring Variable Potential for LLM-based Log Parsing Efficiency and Reduced Costs},
year = {2025},
isbn = {9798400712760},
publisher = {Association for Computing Machinery},
address = {New York, NY, USA},
url = {https://doi.org/10.1145/3696630.3728506},
booktitle = {Proceedings of the 33rd ACM International Conference on the Foundations of Software Engineering},
pages = {596–600},
numpages = {5}
}

\newpage
\appendix

\section{Additional Demonstrations}

\begin{table}[htp]
  \caption{The demonstration of variable-fuzzing strategy.}
  \label{tab:fuzzy}
  \scalebox{0.75}{\begin{tabular}{ll}
      \toprule[1pt]
      \textbf{Raw Log Content} & \textbf{Variable-fuzzed Log Content} \\
      \midrule
      \makecell[l]{"Reopen Block blk\_7008279672769077211"} & \makecell[l]{"Reopen Block blk\_\{Variables\}"} \\
      \addlinespace[0.1cm]
      \makecell[l]{"BLOCK* ask 10.250.14.224:50010 to replicate \\ blk\_-1608999687919862906 to datanode(s) \\ 10.251.215.16:50010 10.251.71.193:50010"}  & \makecell[l]{"BLOCK* ask \{Variable\} to replicate \\ blk\_\{Variable\} to datanode(s) \{Variable\}"} \\
      \bottomrule[1pt]
  \end{tabular}}
\end{table}

\begin{table}[htp]
  \caption{The demonstration of three cache correction scenarios.}
  \label{tab:hulla}
  \scalebox{0.73}{\begin{tabular}{lll}
      \toprule[1pt]
      \textbf{Queried Log Content} & \textbf{LLM Output} & \textbf{Feature} \\
      \midrule
      "i-cache parity error 13" & "i\{-\}cache parity error \{number\}" & punctuation mark \\
      \addlinespace[0.1cm]
      "exception in getting events" & "exception in getting \{label\}" & merely "label" \\
      \addlinespace[0.1cm]
      "invalid uri in request" & "invalid \{uri\} in request" & identical to content \\
      \bottomrule[1pt]
  \end{tabular}}
\end{table}

\begin{figure}[htp]
    \centering
    \includegraphics[width=\linewidth]{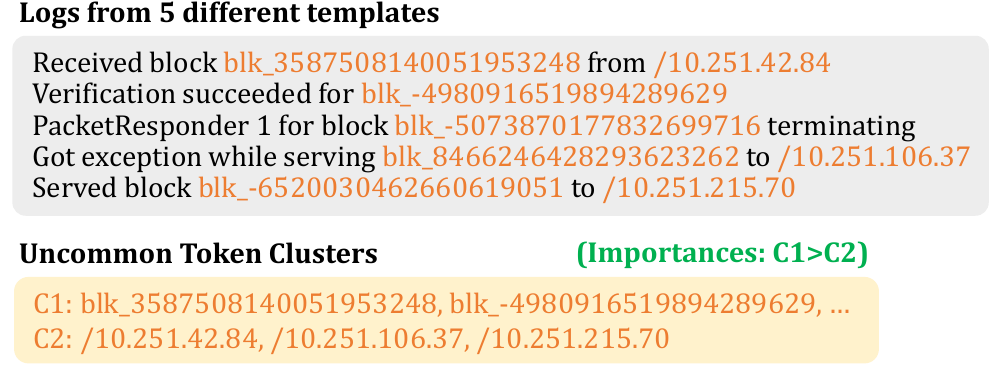}
    \caption{The demonstration of diverse logs and clusters.}
    \label{fig:cluster}
\end{figure}

\begin{figure}[htp]
    \centering
    \includegraphics[width=\linewidth]{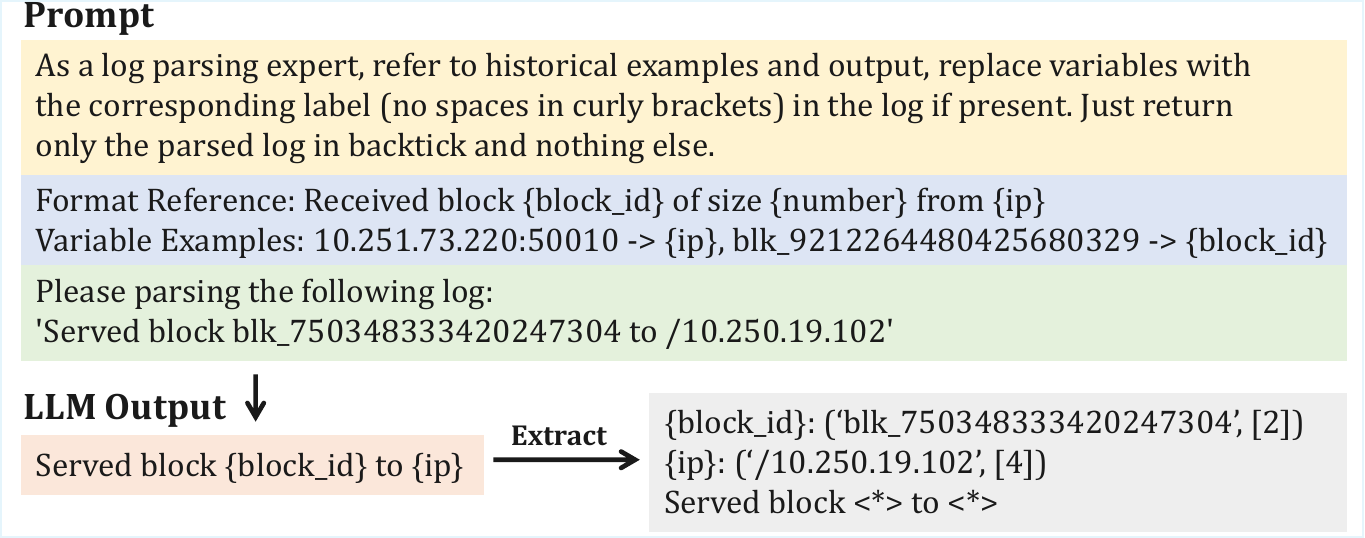}
    \caption{The demonstration of the prompt design.}
    \label{fig:prompt}
\end{figure}

\begin{figure}[H]
    \centering
    \includegraphics[width=\linewidth]{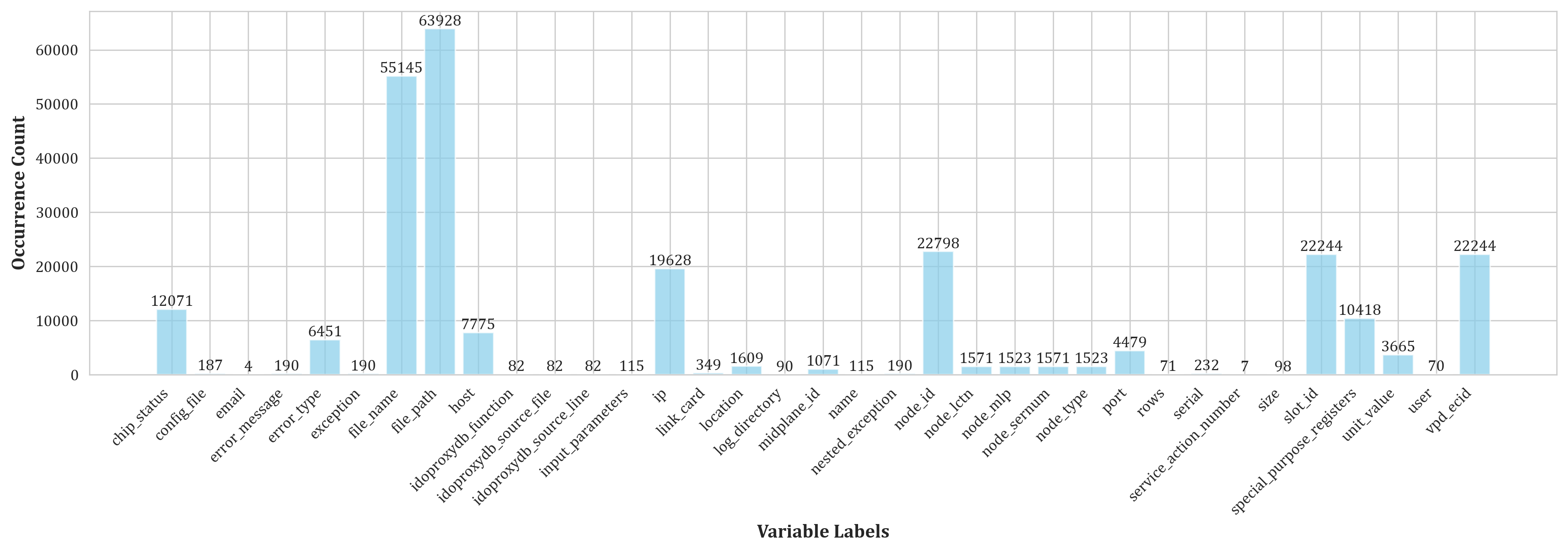}
    \caption{The demonstration of different variable labels and their frequencies in the BGL dataset (For a intuitive analysis, we exclude the label "number").}
    \label{fig:variable}
\end{figure}

\section{Result Integrity of Log Parsing}

Existing log parsing methods typically produce outputs in the form of log templates (constant + placeholder), where variables are replaced with placeholders. This substitution inevitably leads to the loss of original variable information. In practice, however, log variables often carry critical content such as system component states, internal details, and encoded values. Losing this information significantly reduces system observability for engineers and weakens the effectiveness of downstream tasks. 

Following log parsing conventions, our method also outputs log templates (constant + placeholder). Yet, to enhance result quality, we adopt a variable-centric design that preserves richer variable-related information. By storing variable units in cache, our approach captures variable types, specific instances (counts and examples), and their occurrence frequencies across the log set—details entirely discarded in constant-centric designs that retain only placeholders. Since variables represent the dynamic behavior of system modules, this preserved information offers engineers superior visibility into applications and system operations and delivers substantial value for both system analysis and downstream tasks.

\section{Threats to Validity}

We have identified the following major threats to validity.

\textbf{Data Leakage}. Data leakage is a potential concern given that large language models are trained on extensive datasets. For instance, the LLM used in \name might have been trained on open-source log datasets, which theoretically could lead to the memorization of specific templates. However, our experiments reveal that \name's performance in zero-shot scenarios significantly lags behind its performance in few-shot scenarios. This suggests that direct memorization is unlikely. Additionally, in most experiments, \name utilized the gpt-turbo-3.5-0125 model, which was no longer updated before the release of the Loghub-2.0 dataset. Therefore, the possibility of data leakage is negligible. 

\textbf{Privacy Concerns}. Privacy concerns are also paramount for enterprises, as logs often contain detailed customer and service information. Processing internal log data using external LLMs could potentially compromise privacy and security. However, \name is designed as a flexible framework that can integrate various LLMs. Users can integrate their own LLMs into \name, thereby mitigating privacy risks and ensuring data confidentiality.

\section{Evaluation Supplement}

\subsection{Evaluation Baselines}
\label{appd:baselines}

We select eight state-of-the-art baselines, including two syntax-based parsers, two semantic-based parsers, and four LLM-based parsers. In syntax-based parsers, AEL \cite{AEL} uses common formats in log content to capture variables, while Drain \cite{Drain} extracts templates based on prefix tokens and log length. In semantic-based parsers, UniParser \cite{Uniparser} achieves token classification by training a Bi-LSTM \cite{bi-lstm}, while LogPPT \cite{LogPPT} fine-tunes a pre-trained language model (e.g., RoBERTa \cite{roberta}) for log parsing. Among the LLM-based parsers, LILAC \cite{LILAC} helps LLM in parsing by proposing a caching mechanism, while LogBatcher \cite{logbatcher} and LUNAR \cite{Lunar} help LLM in parsing by grouping similar logs, both of which are selected for their leading performance and efficiency. To pursue efficiency and cost-effectiveness, Librelog \cite{LibreLog} uses the small-scale open-source LLM Llama3-8b for log parsing, albeit at the expense of some accuracy in log parsing.

\subsection{Implementation Details}
\label{appd:details}

For semantic-based log parsers that require fine-tuning, we adhere to the community consensus by using their default settings for comparison. In addition, we rigorously follow the configuration of all the LLM-based baselines: using the first 20\% of logs as historical data, setting the default number of candidate logs to 32, and maintaining a consistent Jaccard similarity threshold of 0.8. 

\subsection{Evaluation Metrics}
\label{appd:metrics}

The detailed definitions of evaluation metrics are listed as follows.

\begin{itemize}[leftmargin=*]
    \item Grouping Accuracy (GA): GA is a log-level metric that measures the number of log messages of the same template grouped by the parser. It is computed as the ratio of correctly grouped log messages to the total number of log messages. A log message is considered correctly grouped if and only if its template is aligned with the same set of log messages as the ground truth.
    \item Parsing Accuracy (PA): PA is a log-level metric that measures the ability to correctly extract constants and dynamic variables for each log message. It is computed as the ratio of the number of correctly parsed log messages to the total number of log messages. A log message is considered correctly parsed if and only if all the tokens of templates and variables are correctly recognized.
    \item F1 score of Grouping Accuracy (FGA): FGA is a template-level metric that focuses on the proportion of correctly grouped templates rather than log messages. Specifically, let $N_g$ be the number of templates that are correct in fact and $N_p$ be the number of templates generated by the log parser. If $N_c$ is the number of templates correctly parsed by the log parser, we define grouping accuracy precision (PGA) as $\frac{N_c}{N_p}$  and grouping accuracy recall (RGA) as $\frac{N_c}{N_g}$. Then, FGA is equal to their harmonic mean, which is $\frac{2\times PGA\times RGA}{PGA+RGA}$.
    \item F1 score of Template Accuracy (FTA): FTA is a template-level metric, calculated based on the proportion of correctly recognized templates. It is calculated as the harmonic mean of the precision and recall of the template accuracy. The difference is that a template is considered correctly recognized if and only if the log messages of the parsed template share the same true template and all the tokens of the template are the same as those of the true template.
\end{itemize}

\end{document}